\documentclass[twocolumn]{openjournal}
\usepackage{aas_macros}
\usepackage{color}
\usepackage{soul}
\usepackage{natbib}
\usepackage{hyperref}
\usepackage{amsmath}
\usepackage{xspace}
\usepackage{lineno}
\usepackage{graphicx}
\usepackage{enumitem}
\usepackage{amssymb}
\usepackage{xifthen}
\usepackage{hyperref}
\usepackage[normalem]{ulem}
\usepackage{orcidlink}
\usepackage{longtable}
\usepackage{booktabs} 

\definecolor{vibrantpurple}{RGB}{148, 0, 211}
\definecolor{applegreen}{RGB}{118, 205, 38}
\definecolor{darkred}{RGB}{193, 51, 61}
\definecolor{crimson}{RGB}{220,20,60}

\hypersetup{    
  colorlinks      = {true},
  linkcolor       = {vibrantpurple},
  citecolor       = {vibrantpurple},
  urlcolor        = {vibrantpurple},
}

\newcommand{\unit}[1]{\ensuremath{\mathrm{\,#1}}\xspace}
\newcommand{\feh}{\unit{[Fe/H]}}

\shorttitle{MDF of Ret II}
\shortauthors{Luna et al.}

\begin{document}

\title{A Bimodal Metallicity Distribution Function in the Ultra-Faint Dwarf Galaxy Reticulum II}
\author{Alice~M.~Luna\,\orcidlink{0009-0009-9570-0715}$^{1,2}$}
\author{Alexander~P.~Ji\,\orcidlink{0000-0002-4863-8842}$^{1,2}$}
\author{Anirudh~Chiti\,\orcidlink{0000-0002-7155-679X}$^{1,2}$}
\author{Joshua~D.~Simon\,\orcidlink{0000-0002-4733-4994}$^{3}$}
\author{Daniel~D.~Kelson\,\orcidlink{0000-0003-4727-4327}$^{3}$}
\author{Minsung~Go\,\orcidlink{0009-0008-7277-3818}$^{4}$}
\author{Guilherme~Limberg\,\orcidlink{0000-0002-9269-8287}$^{2}$}
\author{Ting~S.~Li\,\orcidlink{0000-0002-9110-6163}$^{5,6}$}
\author{Anna~Frebel\,\orcidlink{0000-0002-2139-7145}$^{7}$}

\affiliation{$^1$\textit{Department of Astronomy \& Astrophysics, University of Chicago, 5640 S Ellis Avenue, Chicago, IL 60637, USA}}
\affiliation{$^2$\textit{Kavli Institute for Cosmological Physics, University of Chicago, Chicago, IL 60637, USA}}
\affiliation{$^3$\textit{Observatories of the Carnegie Institution for Science, 813 Santa Barbara St., Pasadena, CA 91101, USA}}
\affiliation{$^4$\textit{School of Space Research, Kyung Hee University, 1732 Deogyeong-daero, Yongin-si, Gyeonggi-do 17104, Republic of Korea}}
\affiliation{$^5$\textit{Department of Astrophysical Sciences, Princeton University, Princeton, NJ 08544, USA}}
\affiliation{$^6$\textit{Department of Astronomy and Astrophysics, University of Toronto, 50 St. George Street, Toronto ON, M5S 3H4, Canada}}
\affiliation{$^7$\textit{Department of Physics and Kavli Institute for Astrophysics and Space Research, Massachusetts Institute of Technology, Cambridge, MA 02139, USA}}


\begin{abstract}
Star formation in ultra-faint dwarf galaxies (UFDs, $M_* <10^5M_\odot$) is suppressed by reionization, but may not be completely quenched. The metallicity distribution function (MDF) of stars in ultra-faint dwarf galaxies could show these signatures of reionization. However, past studies of UFD MDFs have been limited, because there are only a few dozen red giant branch (RGB) stars in such low-mass galaxies. We present low-resolution Magellan/IMACS spectroscopy of 167 stars in the UFD Reticulum~II ($M_* \approx  3000  M_\odot$), increasing the number of stellar metallicities by 6.5 times and resulting in the most populated spectroscopic metallicity distribution function of any UFD.  This is possible because we determined the first spectroscopic metallicities of main sequence turn-off stars in any UFD.
The MDF of Reticulum II is clearly a bimodal distribution, displaying two peaks with about $80\%$ of the stars in the metal-poor peak at [Fe/H] $=-3.0$ and $20\%$ of the stars in the more metal-rich peak at [Fe/H] $=-2.1$. Such a large metallicity gap can be explained by Type Ia supernova enrichment during a long quiescent period. This supports the currently-favored two-burst star formation history for Reticulum II and shows that such low-mass galaxies clearly can form stars after reionization. 

\end{abstract}

\section{INTRODUCTION}
\label{intro}
In the $\Lambda$CDM model, massive galaxies such as the Milky Way have many small satellite galaxies due to hierarchical structure formation \citep{Springel2008}. The decades-long ``missing satellite problem" \citep{Klypin1999, Moore1999, Simon&Geha2007, Brooks2013, Kim2018} in which too few satellites were observed was largely resolved as fainter systems were discovered in the era of digital photometric surveys, such as the Sloan Digital Sky Survey (SDSS, \citealt{Willman2005,Zucker2006,Belokurov2008, Belokurov2009, Belokurov2010,Kim2015}) and the Dark Energy Survey (DES, \citealt{Bechtol+2015, Drlica-Wagner2015, Drlica-Wagner2016, Kim&Jerjen2015, Koposov2015a,Koposov2018, Martin2015, Torrealba2016}). The faintest newly discovered satellites became known as ultra-faint dwarf galaxies (UFDs), which are galaxies that live in the smallest dark matter halos and are extremely dark matter dominated \citep{Simon&Geha2007, Willman2012,Simon2019}. They are the most pristine systems, with little to no chemical evolution, due to the majority of their star formation occurring before it is thought that they were quenched during the epoch of reionization \citep{Bullock2000,Benson2002,Somerville2002,Brown2014,RodriguezWimberly2019}. They may also have been important during reionization by contributing ionizing UV photons \citep{Weisz2017, WuKravtsov2024}.

UFDs contain some of the oldest and most metal-poor ($\feh \lesssim -2$) stars since they formed their stars during the early Universe when there were little to no metals in the gas available for star formation \citep{Simon2019}. Therefore, UFDs are relics of the first galaxies formed in the early Universe \citep{Bovill2009, Bovill2011}. This makes them great sites to search for the signatures of the first stars in the Universe, such as enhanced carbon abundances \citep{Frebel&Bromm2012, Frebel&Norris2015, Jeon2017}.

The metallicity distribution function (MDF) is the distribution of individual stellar metallicities within a galaxy. The distribution is determined by a galaxy's star formation history and is affected by stellar yields, feedback, and star formation efficiency. MDFs have been extensively studied in classical dwarf spheroidals (dSphs, $M_* \gtrsim 10^6M_\odot$) such as in \cite{Carigi2002,Lanfranchi2008,Tolstoy2009,Kirby2009,Kirby+2011, Norris2010}. The MDFs are fit with analytic chemical evolution models (e.g., \citealt{Lynden-Bell1975,Pagel1997, Lanfranchi2003, Johnson2021, Sandford2024}) to infer their gas inflows/outflows. These classical dwarf satellites are sufficiently massive that a well-populated MDF can be determined only using metallicities of red giant branch (RGB) stars.  

\begin{deluxetable*}{cccccrrr}
\tablecolumns{8}
\tablecaption{\label{tab:observing}Observations}
\tablehead{Mask & Observing Date &  R.A. & Decl. & $t_\text{exp}$ & Num. of slits & Useful spectra\textcolor{applegreen}{\textbf{*}} & Num. of members \\ 
& & (h:m:s) & (d:m:s) & (s) & & (\%) &} 
\startdata
retIIa & 25 Nov 2019 & 03:35:30 & $-$54:02:50 & 21600 & 126 & 89\% & 94 \\
retIIb & 26 Nov 2019 & 03:35:25 & $-$54:03:00 & 24800 & 122 & 84\% & 82 \\
retIIc & 17 Nov 2023 & 03:35:37 & $-$54:01:50 & 20640 & 99 & 71\%  & 62 
\enddata
\tablecomments{ \textcolor{applegreen}{\textbf{*}} After removing galaxy spectra and SNR $<$ 3.}
\end{deluxetable*}

\begin{figure*}
    \centering
    \includegraphics[width=0.7\linewidth]{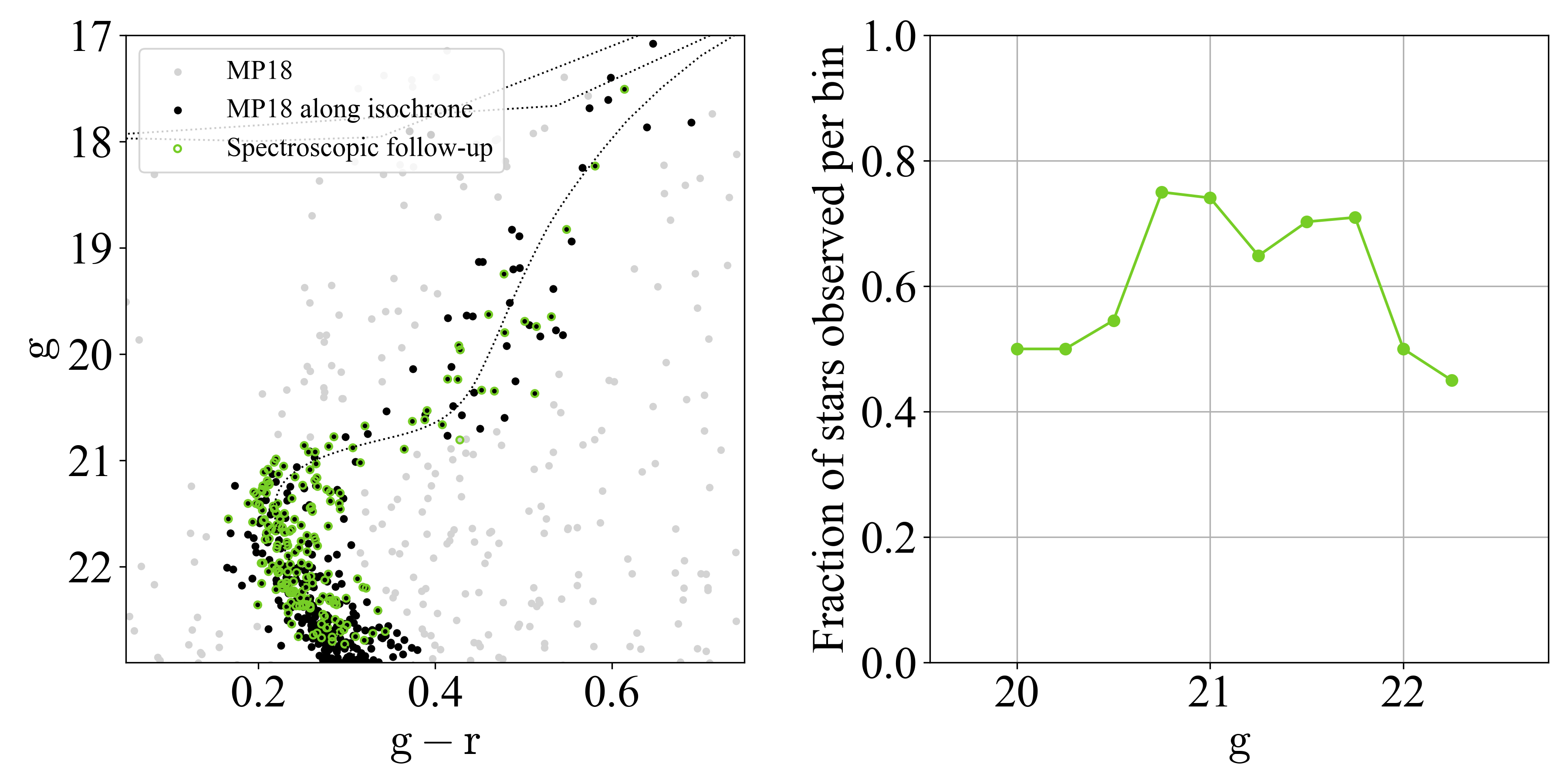}
    \caption{Left: The color-magnitude diagram of the stars in \cite{Mutlu-Pakdil+18} are shown in gray circles. We selected objects along a 13 Gyr isochrone with \feh = $-2$, shown in black circles. The stars we observed for this work are outlined in green. Right: The fraction of stars we observed per 0.5 magnitude bin. The completeness is $\sim$ 70\% for the MSTO stars.}
    \label{fig:MP18completeness}
\end{figure*}

Pushing down to the ultra-faint regime ($M_* <10^5M_\odot$), there are limited metallicity measurements as these galaxies contain fewer RGB stars. \cite{Simon&Geha2007} published the first metallicity measurements in UFDs, but obtaining larger UFD MDF samples still remains an ongoing area of research \citep[e.g.,][]{Norris2010,Lai2011, Vargas2013, Brown2014, Romano2015, Simon+2015,Jenkins2021, Fu+2022, Fu+2023}.
The current best-studied UFD MDF is of Boötes I, which had an MDF of 16 stars immediately after its discovery and soon increased to 41 stars \citep{Norris2010,Lai2011}. A decade later, its most recent MDF consists of 70 stars, the largest spectroscopic MDF of any UFD \citep{Jenkins2021}. Another way to get more populated MDFs is using photometric metallicities, which can be fairly precise by using narrowband filters \citep{Longeard2018,Chiti2020,Fu+2022,Fu+2023}. With deep narrowband CaHK HST photometry, \cite{Fu+2022} obtained metallicities of 60 RGB stars in Eridanus II, noting that their results are consistent with spectroscopic studies \citep{Li2017}, but with a larger scatter.

In this paper, we study the MDF of Reticulum II (Ret II), an ultra-faint dwarf galaxy with stellar mass of $M_* = 10^{3.51\pm0.04} M_\odot$ \citep{Mutlu-Pakdil+18}. Ret II was discovered from the first data release of the Dark Energy Survey (DES) in 2015 \citep{Bechtol+2015, Koposov2015a} and quickly confirmed to be a galaxy \citep{Simon+2015,Walker2015,Koposov2015b}. As one of the closest UFDs discovered at only 32 kpc, it became a galaxy of interest due to the feasibility of deeper studies. It was initially speculated to have a connection to the Magellanic Clouds (MCs), which was later confirmed with orbital histories \citep{Patel2020,Erkal2020}. It was also of interest because of the tentative detection of gamma-ray emission (e.g., \citealt{GeringerSameth2015}), which could be due to dark matter particles self-annihilating into high-energy gamma rays. 
With high-resolution follow-up spectroscopy, Ret II was the first UFD discovered to have been enriched by a rare and prolific r-process nucleosynthesis event, either a neutron star merger or a rare core-collapse supernova \citep{Ji+2016a, Roederer2016}.
More recently, several works have studied the detailed formation history of Ret II. The star formation history (SFH) described in \cite{Simon+2023} favors a two-burst star formation scenario. \cite{Ji+2023} details the low dispersion in the r-process enriched stars. Lastly, \cite{Fu+2023} obtained a photometric MDF from 76 metallicities (50 constrained and 26 upper limits).

Ret II has only 20 stars with spectroscopic metallicities in the literature, excluding upper limits \citep{Simon+2015, Ji+2016a, Ji+2023}. This paper aims to determine a well-populated spectroscopic MDF. Due to the small number of RGB stars, we had to determine metallicities of the faint main sequence turn-off (MSTO) stars (g $\lesssim$ 22), the first time this has been done in any dwarf spheroidal fainter than the Magellanic Clouds.

In Section \ref{obs}, we outline the data and observations. In Section \ref{methods}, we describe how we determine metallicities and compare to the literature. In Section \ref{analysis}, we present the MDF and investigate different analytical and chemical evolution models. We discuss implications for Ret II's formation history in Section \ref{discussion} and conclude in Section \ref{conclusion}.

\begin{figure*}
    \centering
    \includegraphics[width=\linewidth]{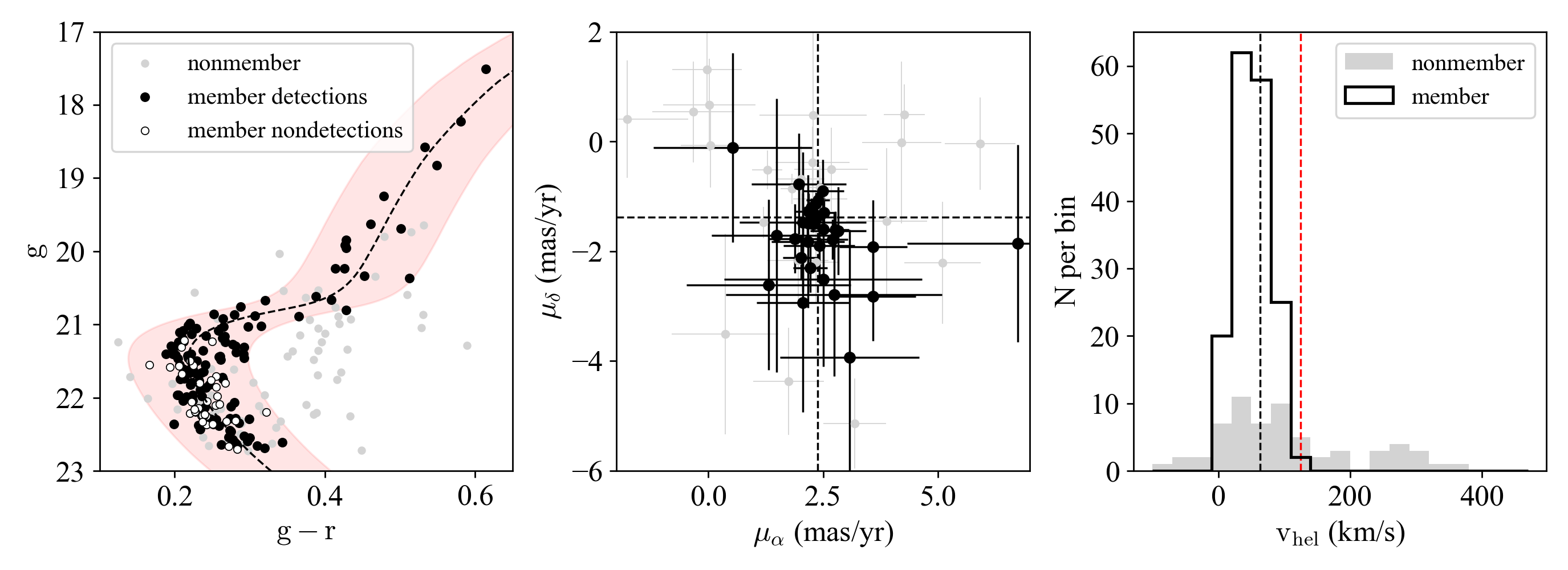}
    \caption{Membership selection criteria. {Left:} Color-magnitude diagram of Reticulum II using \citet{Mutlu-Pakdil+18} photometry. We overlaid a MIST isochrone of \feh =  $-2$ and age = 13 Gyr. The red filled region is $\Delta\delta(g-r) \pm 0.08$ from the isochrone. Stars with $\Delta (g-r) > 0.08$ from the isochrone were rejected. Detection means that a metallicity was measured from a detected Ca II K line. Nondetections have 3$\sigma$ upper limits. 
    {Center:} Proper motions for the subset of stars in Gaia DR3. The black dashed lines ($\mu_\alpha = 2.38$ mas yr$^{-1}$ and $\mu_\delta = -1.38$ mas yr$^{-1}$ ) are the Gaia mean proper motions of Ret II from \cite{Pace2022}. Stars with proper motions $> 3\sigma$ away from that of Ret II were rejected. There were no member nondetections (open circles) with proper motions.
    {Right:} Velocity distribution of the observed sample. The member histogram includes detections and nondetections. The black dashed line is the mean velocity of Ret II $\text{v}_\text{hel} = 63.9$  km s$^{-1}$ from \cite{Ji+2023} and the red dashed line is $\text{v}_\text{hel} = 125$  km s$^{-1}$. Stars with $\text{v}_\text{hel}>$ 125 km s$^{-1}$ were rejected. The nonmembers with v $<$ 125 km s$^{-1}$ are removed by other cuts. Nonmember stars in each panel are those rejected by combining all three cuts.}
    \label{fig:diagnosticplots}
\end{figure*}

\section{OBSERVATIONS}
\label{obs}
We observed stars in Reticulum II with the Inamori-Magellan Areal Camera and Spectrograph (IMACS; \citealt{Dressler+2011}) with the f/2 camera on the Magellan Baade telescope at Las Campanas Observatory on 2019 November 26 and 27 with follow-up observations on 2023 November 17. 
The targets were selected in 3 ways: (1) along the isochrones of the deep ($g \sim 25$) photometric catalogs provided by \citet{Mutlu-Pakdil+18} using Megacam \citep{McLeod2015} on the Magellan Clay Telescope, (2) similarly from \citet{Simon+2023} using Advanced Camera for Surveys (ACS; \citealt{Ford+2003}) on the Hubble Space Telescope, and (3) from a sample of proper motion members in \citet{Pace&Li+19} using DES DR1 \citep{Bechtol+2015,Drlica-Wagner2018,Abbott2018} crossmatched with Gaia DR2 \citep{GaiaCollab+2018a:Brown}. We obtained low-resolution spectroscopy (R$\sim$1500 near the blue end of the spectrum) using the 400 $l$/mm grism with a central wavelength of $4730$\AA. The 3 observing nights described in Table \ref{tab:observing} each used different multi-slit masks with 1” slits, totaling 245 unique objects. 
The wavelength range for each spectrum spans $3700-5800$\AA. Each slitmask had $\sim$ 6 hours of total exposure time allowing us to reach signal-to-noise (SNR) $\sim$ 10 per 0.9\AA \ pixel near CaK (3933\AA) for stars near the MSTO of Reticulum II at \textit{g} $\sim$ 21.
In Figure \ref{fig:MP18completeness}, the left panel shows the CMD of stars in the \cite{Mutlu-Pakdil+18} catalog in gray circles and the stars selected for spectroscopic follow-up in green circles, and the right panel shows the fraction of photometric candidates from \cite{Mutlu-Pakdil+18} observed per magnitude bin. For the MSTO stars, we have $\sim 70\%$ completeness.

We followed \citet{Newman+2020} using the CarPy \citep{Kelson+2000,Kelson2003} pipeline for data reduction. The following procedure is run per slitmask. A mask image is taken during the afternoon that maps the positions of the slits to observed objects. Each science frame has a corresponding HeHg arc lamp spectrum for wavelength calibration and finding the object position. The pipeline uses the [OI] 5577$\text{\AA}$ sky emission line to tweak the wavelength solution before sky subtraction. Flat-field twilight spectra are used to divide out pixel-to-pixel variations of the eight CCD detectors. CarPy provides the 2D wavelength calibrated rectified spectra. The one-dimensional spectrum is then extracted using the optimal method described in \cite{Horne1986} that traces the stellar profile to apply larger weights to pixels that contribute more flux. On the final observing night, issues with the Atmospheric Dispersion Corrector (ADC) resulted in distortions in the spatial direction. We thus added an additional correction by fitting out this spatial distortion.

\section{METHODS}
\label{methods}
\subsection{Membership selection}
\begin{figure*}
    \centering
    \includegraphics[width=0.75\linewidth]{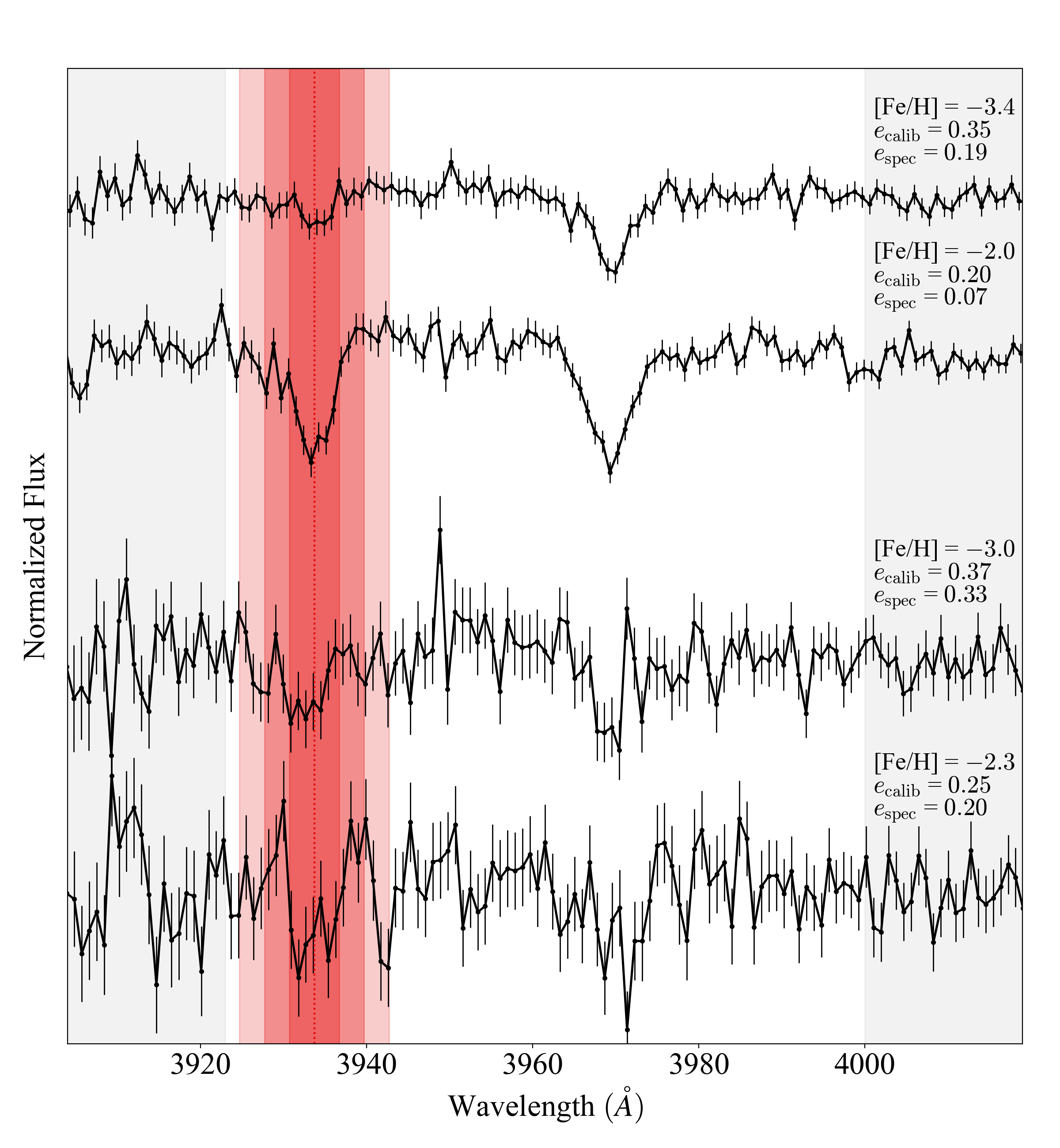}
    \caption{Example spectra with similar $B-V$ around the Calcium HK region. The red region is centered on the Ca II K line at 3933.7 \AA \ with different bands of the KP Index. The grey regions on either side of the spectrum are used for local continuum normalization. The top two spectra have SNR $\sim $ 11.5 and the bottom two have SNR $\sim $ 4.5. The calibration uncertainty, $e_\text{calib}$, is the largest source of uncertainty and dominates 75\% of our stars. The spectroscopic uncertainty, $e_\text{spec}$, decreases with higher metallicity and SNR. \\}
    \label{fig:spectrum}
\end{figure*}
We observed spectra of 245 objects from photometric catalogs of Ret II in the literature. After cleaning out 7 clear galaxy/quasar contaminants and 7 other non-stellar spectra, we made further membership selections to reject possible Milky Way contaminants. We required the data to have SNR $>$ 3 to ensure good data quality. Some stars were observed over multiple nights. For these stars, the total flux of a star is the sum of the fluxes of each night, and the total uncertainty is the quadrature sum of the uncertainties. We selected members along a 13 Gyr isochrone with $\feh = -2$ using MESA Isochrones and Stellar Tracks (MIST; \citealt{Dotter2016,Choi2016,Paxton2011, Paxton2013, Paxton2015}) and \texttt{Minimint} (MIni Mist INTerpolation; \citealt{sergey_koposov_2023_10437032}).
We used a distance modulus of $\mu =$ 17.5 and reddening of $E(B-V)=$ 0.04, within the uncertainty of 0.01 mag measured in \cite{Simon+2023}. We rejected 28 outliers with a distance from the isochrone of $\Delta(g-r)>$ 0.08. We had a subset of 50 stars with proper motions from Gaia DR3, and rejected 8 stars larger than 3$\sigma$ from the mean proper motion of Ret II found in \cite{Pace2022}. Figure \ref{fig:diagnosticplots} shows the membership selection criteria. We also rejected 7 stars that \cite{Simon+2015} labeled as nonmembers based on their velocities. 

We found radial velocities by cross-correlating the stellar spectra with the high-resolution rest-frame template spectrum of HD21581 around the H$\gamma$ (4340\AA) and H$\beta$ (4861\AA) lines. HD21581 is an RGB star with a heliocentric radial velocity of 153.7 km s$^{-1}$ and $\feh = -1.82$, a higher metallicity than the average metallicity in Reticulum II \citep{Roederer2014}. The overlapping wavelength range between our spectra and HD21581 is 3800 to 5550\AA. We used the strong absorption lines near the redder end of the spectrum to avoid the lines at the noisier blue end and minimize atmospheric dispersion effects. Only one star (HST\_037) did not have a velocity because the H$\gamma$ and H$\beta$ lines were near the chip gaps in the detector, but it passed the velocity cut with the Ca K line velocity. We show the velocity histogram in Figure \ref{fig:diagnosticplots}. We rejected 21 stars with velocities larger than 125 km s$^{-1}$, too large to be consistent with the mean velocity of Reticulum II. Given the low resolution, the velocity uncertainties are $>10 $ km s$^{-1}$, which is not useful for the internal kinematics of the galaxy. Implementing all membership selection criteria resulted in a sample of 167 members.
\begin{figure}
    \centering
    \includegraphics[width=\linewidth]{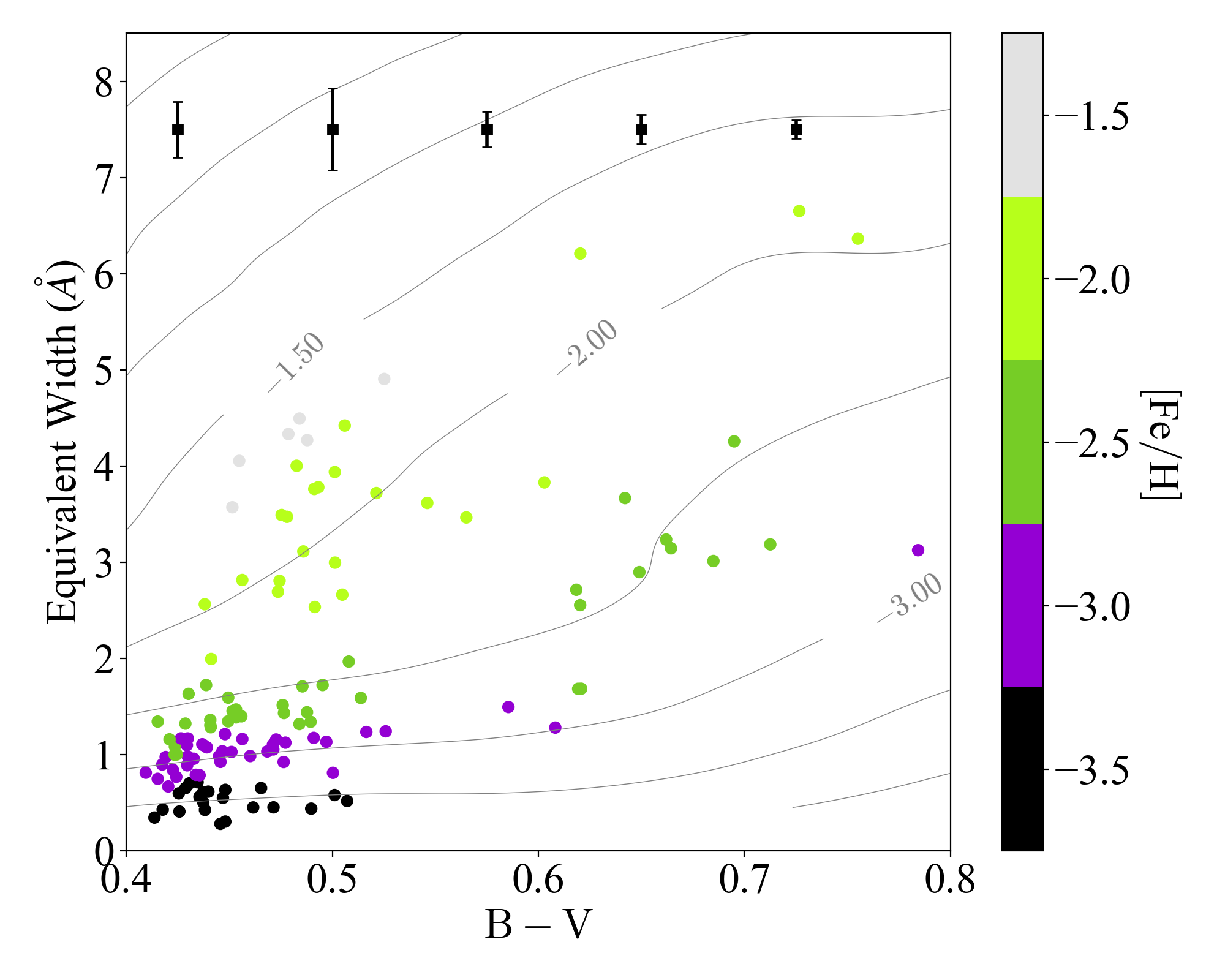}
    \caption{The equivalent width (EW) and $B-V$ color index used to get metallicity for member stars with detected Ca K lines. Metallicity increases with a larger EW and lower $B-V$. The gray contour lines are constant metallicity. The typical EW uncertainties are shown at the top in black squares for 0.1 magnitude bins. At $B-V \lesssim 0.5$, we see a clear gap between stars with \feh = $-2$ (bright green) and stars with \feh = $-2.5$ (dark green), indicating possible bimodality in \feh. A discrete colorbar was used to emphasize the gap in metallicity.}
    \label{fig:kp_bv}
\end{figure} 
\subsection{Measuring [Fe/H]}
\label{measure_feh}
We measured metallicities from a calibration \citep{Beers+1999} using a pseudo-equivalent width (EW) of the Ca II K line (3933\AA) and $B-V$ color index as input parameters. The EW measurement followed the KP Index method described in the calibration paper and outlined in \cite{Chiti2018}. We calculated the area under the Ca II K line within 6\AA, 12\AA, and 18\AA \ bands centered at 3933.7\AA. Figure \ref{fig:spectrum} shows the spectral range used for the analysis (K6 from $3930.7-3936.7$\AA, K12 from $3927.7-3939.7$\AA, and K18 from $3924.7-3942.7$\AA). The resulting equivalent width adopts the KP index value based on the following criteria:
\[
\text{KP} =
\begin{cases} 
\text{K6} & \text{if } \text{K6} \leq 2 \, \text{\AA}, \\
\text{K12} & \text{if } \text{K6} > 2 \, \text{\AA} \text{ and } \text{K12} \leq 5 \, \text{\AA}, \\
\text{K18} & \text{if } \text{K12} > 5 \, \text{\AA}.
\end{cases}
\]
The regions from $3903-3920$\AA \ and $4000-4020$\AA \ are used for local continuum normalization. We visually inspected each spectrum and adjusted the normalization region as needed due to chip gaps. The EW uncertainties were calculated using the following equation, 
\begin{equation}
    \sigma_\text{EW} = \Delta\lambda \ \cdot \sqrt{\sum_i^K \sigma_i^2},
\end{equation}
where $\Delta\lambda$ is the dispersion of the 400 l/mm grism $\sim 0.9$\AA \ per pixel, $\sigma_i$ is the normalized flux uncertainity at wavelength, $i$, and $K$ is the length of the KP region. We report the EW and $\sigma_\text{EW}$ values in Appendix \ref{metallicity_tab_appendix}.
We fit a Gaussian profile to the Ca II K line for each star to check for line detections. The line fit for each star was inspected by eye to confirm whether there was a real detection. We determined 129 detections and 38 non-detections. We show the EW and $B-V$ values colored by metallicity in Figure \ref{fig:kp_bv}. The typical EW uncertainties are 0.29, 0.43, 0.19, 0.15, and 0.09\AA \ for the following $B-V$ color bins: $0.40-0.45, 0.45-0.55, 0.55-0.6, 0.6-0.7,\ \text{and}\ 0.7-0.8$. For non-detections, we calculated $3 \sigma_\text{EW}$ upper limits. The $\sigma_\text{EW}$ are propagated into the metallicity uncertainties, $e_\text{spec}$.
The majority of stars were dominated by the systematic uncertainty in the calibration, $e_\text{calib}$. In \cite{Beers+1999}, there were not many stars in their sample near the low KP and low $B-V$ regions, leading to uncertainties of $e_\text{calib} \sim 0.2-0.4$. The color does not appear to have an effect on the systematic uncertainties, while higher equivalent widths have lower uncertainties.

\begin{figure}
    \centering
    \includegraphics[width=\linewidth]{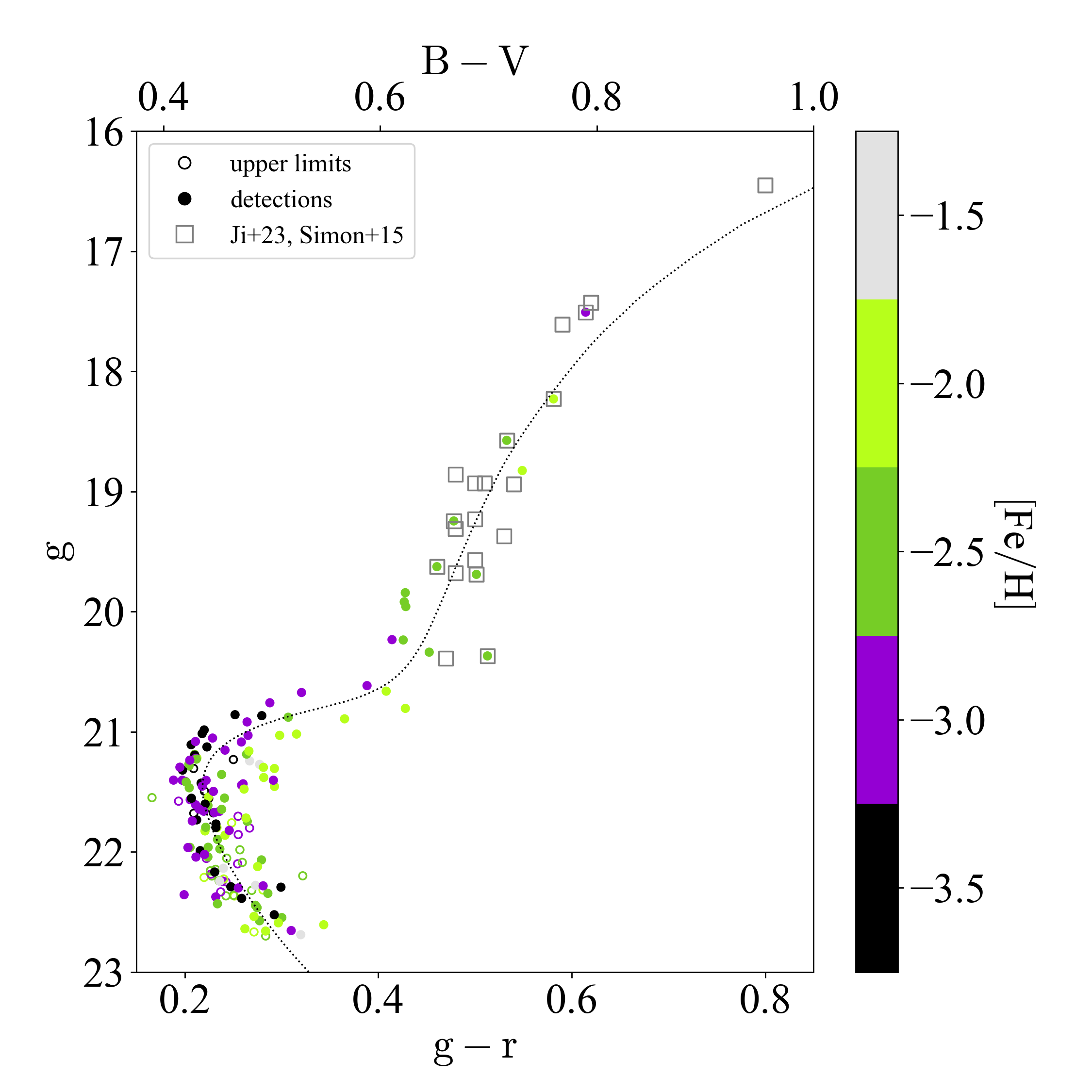}
    \caption{The color-magnitude diagram of members in Reticulum II. We included a 13 Gyr MIST isochrone with [Fe/H]= $-2$. All previous spectroscopic observations are in squares. Stars with detected Ca II K lines have measured [Fe/H] abundances in closed circles. Stars with upper limit metallicities are in open circles. This diagram emphasizes (1) how faint we were able to measure metallicities compared to what we currently have in the literature and (2) that we have observations of the main sequence for the first time in Reticulum II. \\ } 
    \label{fig:cmd}
\end{figure}

We used Magellan/Megacam photometry from \citet{Mutlu-Pakdil+18} for the $B-V$ input. After dereddening, $g$ and $r$-band magnitudes were transformed to SDSS magnitudes \citep{Drlica-Wagner2018} then to UBVRI \citep{Jordi+2006}. Figure \ref{fig:cmd} shows the color-magnitude diagram of the member stars colored by their metallicities. We observed 10 stars without Magellan/Megacam photometry. From 150 stars with Magellan/Megacam, DES, and HST magnitudes, we compared the magnitude systems and determined the coefficients for each magnitude relation. We used this to transform DES and HST magnitudes onto a scale similar to the Magellan/Megacam magnitudes. The scatter of the relation ($\sigma \sim 0.2$ mag) was propagated in the photometric uncertainty, $e_\text{phot}$, for the 10 stars. The total metallicity uncertainty for a star is $e_\text{total} = \sqrt{e_{\text{spec}}^2 + e_{\text{calib}}^2 + e_{\text{phot}}^2}$. 

\subsubsection{Zero-point calibration}
\label{comparing}
\begin{figure}
    \centering
    \includegraphics[width=\linewidth]{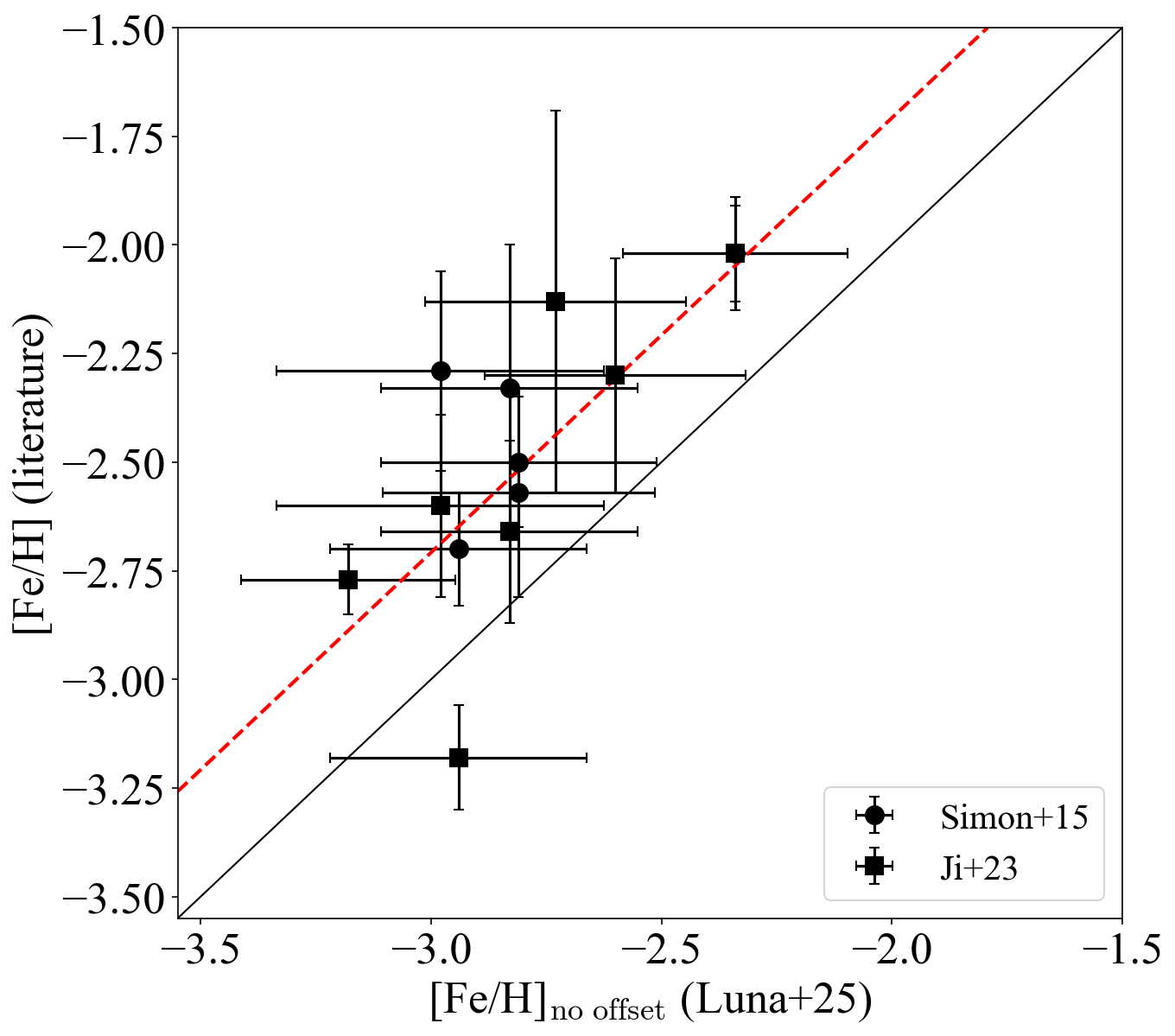}
    \caption{\feh values measured in this work compared to \cite{Simon+2015} and \cite{Ji+2023}. The red dashed line shows an offset of 0.29 dex compared to the one-to-one line in black. There are 9 stars in our sample found in the literature, with 4 of them having multiple measurements and resulting in 13 data points used in the comparison. An offset of 0.29 was then added to all metallicities and upper limit values.} 
    \label{fig:comparison}
\end{figure}
It is not clear if the Ca II K calibration \citep{Beers+1999} is on the same scale as most of the literature on dwarf galaxies that uses the Calcium II triplet (CaT) calibration \citep{Carrera2013}. For our sample, the CaT calibration is not ideal due to weaker lines in hotter MSTO stars. Additionally, there is no published CaT calibration for MS stars. We compared our metallicities to medium-resolution spectroscopic metallicities with VLT/FLAMES from \citet{Simon+2015} and \citet{Ji+2023}. \cite{Simon+2015} used the CaT lines (8498, 8542, 8662\AA) and \cite{Ji+2023} used an Fe I line (6495\AA) to measure \feh. We have 9 stars cross-matched with \cite{Simon+2015} and \cite{Ji+2023}, with 4 stars observed in both. The 13 metallicities used for our comparison are shown in Figure \ref{fig:comparison}. We note that \cite{Ji+2023} uses Dartmouth isochrones \citep{Dotter2008}, so metallicities may not be directly comparable. There were 12 measurements more metal-poor and 1 measurement more metal-rich than in the literature, with a weighted average offset of $-0.29$ dex. As literature values mostly use CaT spectroscopic metallicities, we apply a $+$0.29 dex shift to all of our reported \feh values, including upper limits. Future work will focus on a full recalibration of this metallicity scale.

\section{ANALYSIS}
\label{analysis}

\subsection{MDF of RetII}
\label{mdf_comparing}
\begin{figure}
    \centering
    \includegraphics[width=\linewidth]{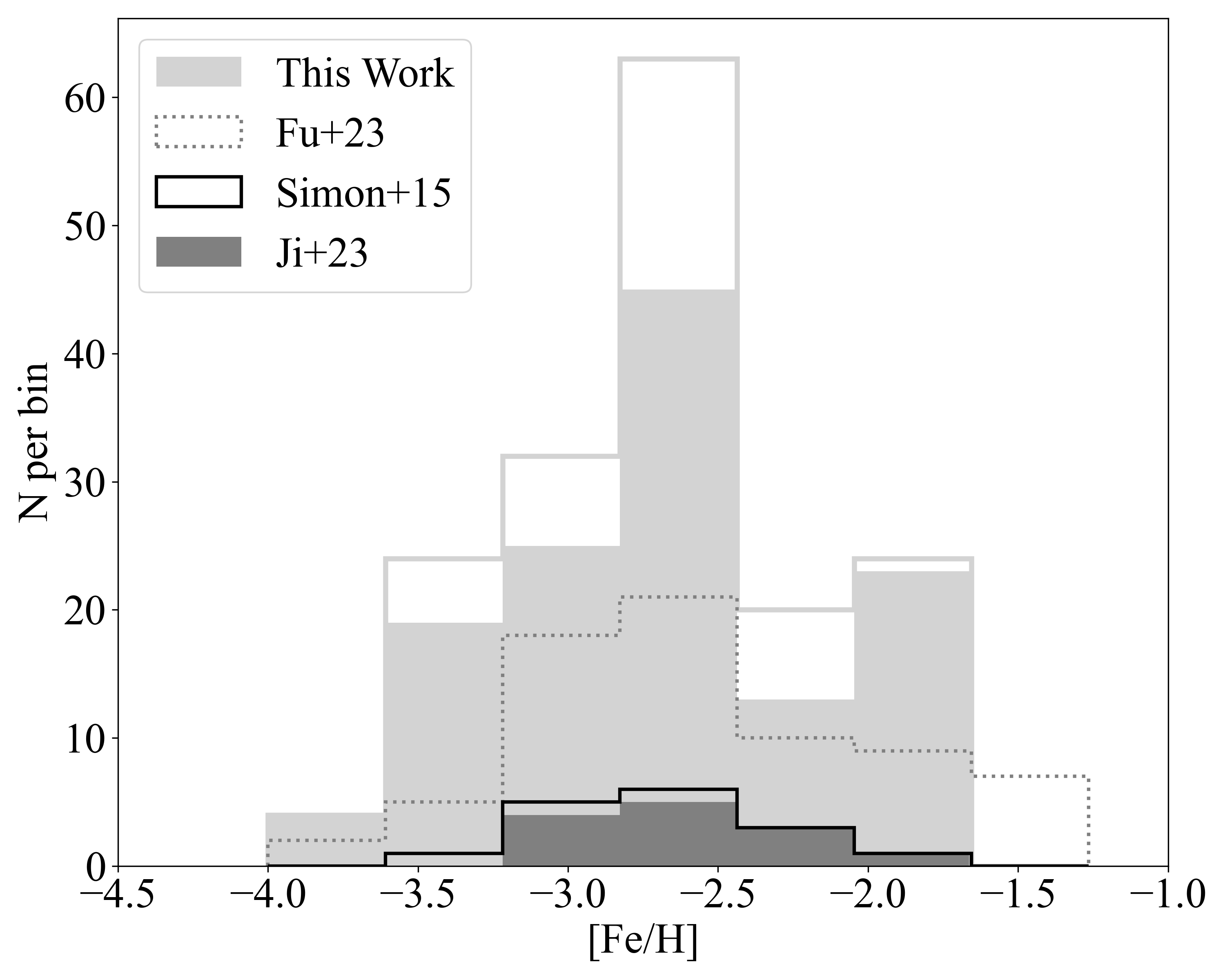}
    \caption{The metallicity distribution function of Reticulum II with N=129 spectroscopic metallicities (gray histogram) and N=38 upper limits (gray line stacked on gray histogram). Previous work \cite{Simon+2015} with N=16, \cite{Ji+2023} with N=13, and \cite{Fu+2023} N=76 (50 constrained). Binned by the median $e_\text{total}$ $\sim $ 0.4.}
    \label{fig:mdf_lit}
\end{figure}
\begin{figure}
    \centering
    \includegraphics[width=\linewidth]{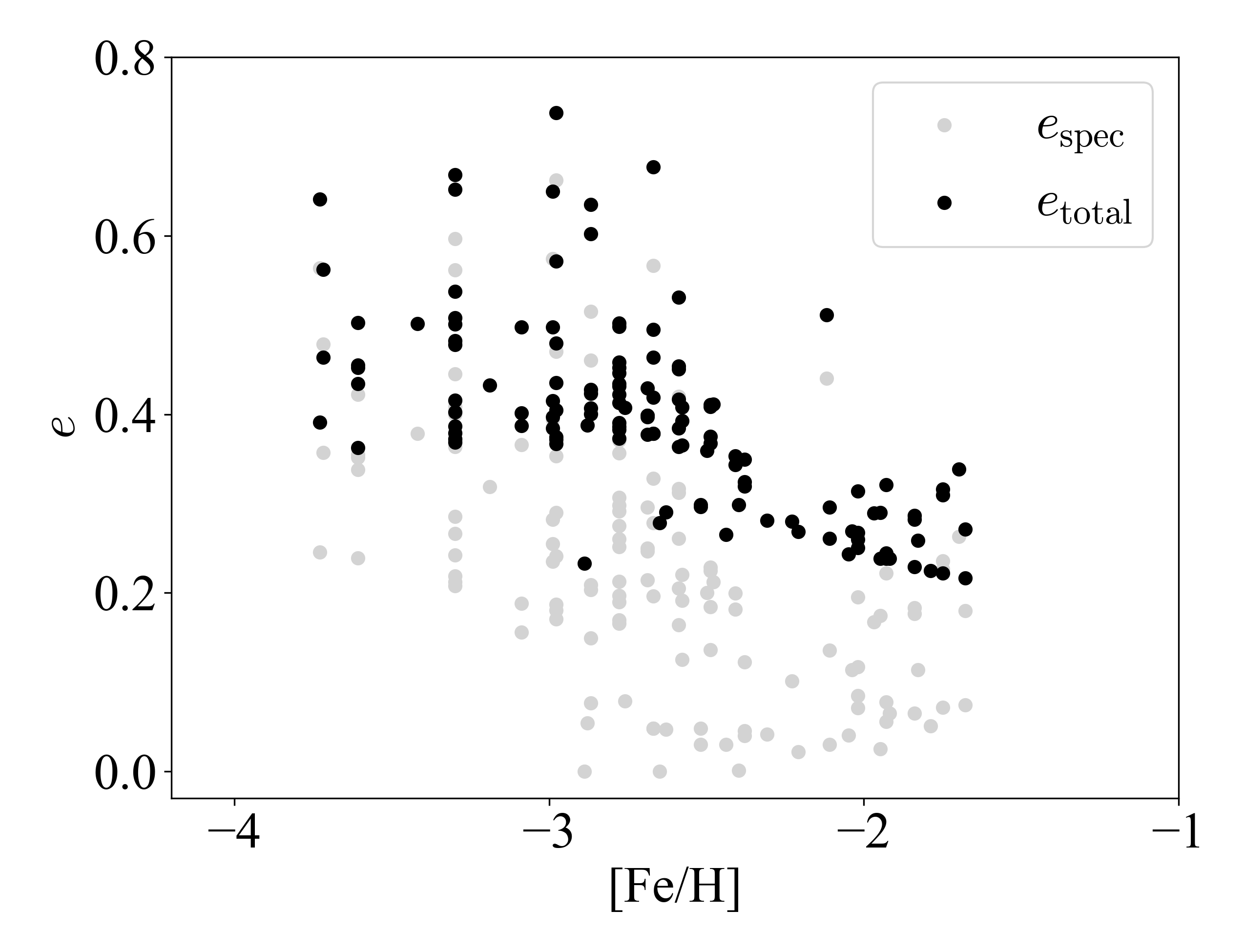}
    \caption{The metallicity and metallicity uncertainties for stars with detections. The stars at \feh$\gtrsim -2.5$ have lower uncertainties, making us more confident in the bimodality of the  MDF. Lower metallicity stars have larger uncertainties for many reasons, one being larger systematic uncertainties as seen in Figure \ref{fig:spectrum}.}
    \label{fig:metallicity_diff}
\end{figure}

In Figure \ref{fig:mdf_lit}, we show the metallicity distribution function (MDF) of Reticulum II with 129 metallicities in the light gray filled histogram and the 38 upper limits stacked on top in the light gray line. Our MDF is compared to literature spectroscopic \citep{Simon+2015, Ji+2023} and photometric \citep{Fu+2023} MDFs. We binned the histogram by the median metallicity uncertainty of 0.4 dex. The number of stars with spectroscopic metallicities in Ret II increased by $\sim$ 6.5x, making this the most populated spectroscopic or photometric UFD MDF currently in the literature. This was possible due to our spectroscopic metallicities of faint MSTO stars. By eye, we see a bimodality in the MDF that has not been seen before in any UFD, possibly because of the small number of stars observed. At \feh $\gtrsim -2.5$, the uncertainties are lower by 0.2 dex, further motivating the investigation of bimodality, see Figure \ref{fig:metallicity_diff}.

\begin{table}
\centering
\caption{\label{tab:gaussian_literature} Literature Gaussian MDF Parameters}
\begin{tabular}{l||cccc}
\toprule
\colhead{Reference} & \colhead{$\langle \text{[Fe/H]} \rangle$} & \colhead{$\sigma$}& \colhead{$N_{\text{total}}$} & \colhead{$N_{\text{lim}}$} \\
\midrule
Simon et al. 2015 & $-2.65_{-0.07}^{+0.07}$ & 0.28$_{-0.09}^{+0.09}$ & 16  & 0 \\
Ji et al. 2023 & $-2.64_{-0.11}^{+0.11}$ & $0.32_{-0.07}^{+0.10}$ & 13 & 0 \\
\hline
Luna et al. 2025 \\
\ \ \ {This work, All} & $-2.78_{-0.05}^{+0.05}$ & $0.50_{-0.04}^{+0.04}$ & 167 & 38 \\
\ \ \ {This work, Det. only} & $-2.59_{-0.05}^{+0.05}$ & $0.40_{-0.05}^{+0.05}$ & 129 & 0 \\
\hline
Fu et al. 2023 & $-2.64^{+0.10}_{-0.11}$ & $0.72^{+0.09}_{-0.08}$ & 76 & 26 \\
\bottomrule
\end{tabular}
\end{table}

\subsection{Likelihood Analysis of MDF}

We follow \cite{Kirby+2011,Kirby2013} for the likelihood functions of the detections and \cite{Ji+2023} for the upper limits. We determined the best-fit model parameters, $\theta$, by maximizing the likelihood function, $L$, or the more computationally tractable optimization of the loglikelihood function, ${\mathcal{L} = \ln (L) = \ln (\prod_i L_i)= \sum_{i}\ln L_i}$, where $L_i$ is the likelihood for one star, $i$.

For detections, we adopt a Gaussian likelihood,
\begin{equation}
    \label{star_gaussian}
    P_{\text{det},i}(z\mid\feh_i, e_i)= \frac{1}{\sqrt{2\pi}e_i} \exp\frac{-(z-\feh_i)^2}{2e_i^2},
\end{equation}
with $z$ as the possible range of metallicities, $\feh_i$ as the metallicity of a star, and $e_i$ as the metallicity uncertainty of a star.
The likelihood for the detections is
\begin{equation}
\label{detection_likelihood}
{L_{\text{det},i}  =  \int_{-\infty}^{\infty} P_{\text{det},i}(z\mid\feh_i, e_i) \cdot P_{\text{M}}(z\mid\theta) \ dz,}
\end{equation}
where $P_{\text{M}}$ is the probability distribution for different models.

We use the upper limit metallicities, $\feh_{\text{lim}}$, as an additional constraint. To include the 3$\sigma$ upper limits, we added a likelihood function such that there is a 99.7\% chance that $z < \feh_{\text{lim}}$ and a 0.3\% chance that $z > \feh_{\text{lim}}$. We adopt a step-function described as $P_{\text{lim},i}(z\mid\feh_{\text{lim},i}): P_M(z\leq \feh_{\text{lim}, i}) = 0.997$. We use the cumulative distribution function given the best-fit model parameters, $F(z\mid\theta)$, and the complementary cumulative distribution function, $\bar{F}(z) = 1-F(z)$. The likelihood for the upper limits is
\begin{multline}
\label{limit_likelihood}
{L_{\text{lim},i}  =  \int_{-\infty}^{\infty} P_{\text{lim},i}(z\mid\feh_{\text{lim},i}) \cdot P_{\text{M}}(z\mid\theta) \ dz}
\end{multline}
\begin{multline}
    \ \ \ \ \ \ \ = 0.997 \cdot F(\feh_{\text{lim},i}\mid\theta) \\ + 0.003 \cdot \bar F(\feh_{\text{lim},i}\mid\theta).
\end{multline}
The total loglikelihood is the sum of the loglikelihoods of the detections and upper limits, $\mathcal{L}_{tot} = \sum_i \ln{L}_{\text{det},i} + \sum_i\ln{L}_{\text{lim},i}$.

We sampled the posterior with \texttt{emcee} \citep{emcee2013} and initialized 64 walkers with 2000 burn-in iterations. We assumed log-flat priors on the parameters to limit the metallicity range explored. To determine a goodness of fit for each model, we use the maximum likelihood from the best-fit parameters and penalize any overfitting with extra parameters. We adopt the Akaike Information Criterion (AIC) \citep{Akaike1974} described as:
\begin{equation}
    \text{AIC} = 2k  -2 \ln (L), 
\end{equation}
where $k$ is the number of parameters in the model. A more complex model is likely to show a better fit, but to account for overfitting, there is a penalization of extra parameters ($+2k$). A lower AIC value indicates a better fit. 
\subsection{Single Gaussian MDF}
We determine the model parameters for a single Gaussian MDF, as previously done in the literature (e.g., \citealt{Li2018}). 
The Gaussian MDF has two parameters, mean metallicity, $\langle \text{[Fe/H]} \rangle$, and metallicity dispersion, $\sigma$:
\begin{multline}
    P_\text{M} (z\mid \langle \text{[Fe/H]} \rangle, \sigma) = \frac{1}{\sqrt{2\pi}\sigma}\exp\frac{-(z-\langle \text{[Fe/H]} \rangle)^2}{2\sigma^2}
\end{multline}

We use equations \ref{detection_likelihood} and \ref{limit_likelihood} to maximize the total likelihood of detections and upper limits with the Gaussian model. For the 167 stars, we get a mean metallicity, $\langle \text{[Fe/H]} \rangle = - 2.78 \pm0.05$, and dispersion, $\sigma$ = 0.50$\pm$ 0.04. We note that when we only include detections (N=129), the mean metallicity and dispersion are more consistent with the literature, with $\langle \text{[Fe/H]} \rangle$ = $-2.59\pm$ 0.05, and a dispersion, $\sigma = 0.40\pm$ 0.05.

Table \ref{tab:gaussian_literature} shows the mean metallicities and dispersions of spectroscopic and photometric studies. The MDF in \cite{Simon+2015} was determined from 16 spectroscopic metallicities with the Calcium triplet calibration of \cite{Carrera2013}. The MDF in \cite{Ji+2023} is from 13 stars with an iron line fit using the radiative transfer and spectral synthesis code MOOG \citep{Sneden1973}. \cite{Fu+2023} measured the metallicities for Ret II stars along the MSTO and main sequence (MS) using HST narrowband CaHK imaging and bolometric corrections from MIST. Their MDF has 76 photometric metallicities (including 26 upper limits), with a larger dispersion compared to spectroscopic studies. Their sample contains 1/3 of stars with $\feh > -2$, unlike spectroscopic MDFs, including ours. 
\subsection{Analytical Chemical Evolution Models}
\begin{figure}
    \centering
    \includegraphics[width=\linewidth]{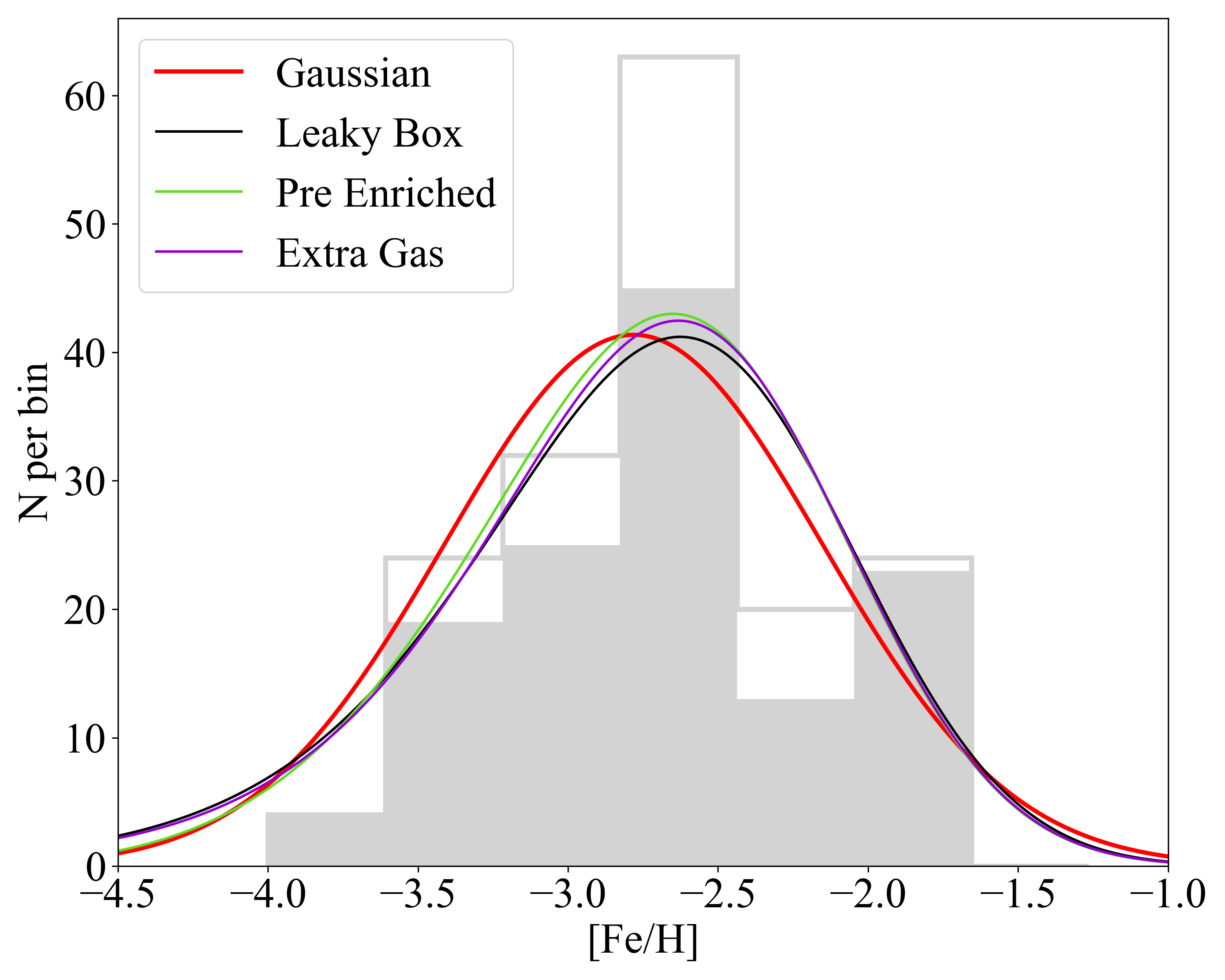}
    \caption{Analytical Chemical Evolution models and Gaussian MDF convolved with the typical uncertainty of $\sim 0.4$ dex. The Gaussian model has the lowest AIC and the best fit. The Extra Gas model is a worse fit. A more complex model is necessary to provide a better fit to the data.}
    \label{fig:gce_gauss_plot}
\end{figure}
\begin{table}
\centering
\caption{\label{tab:gce_params} Chemical Evolution Model Parameters}
\begin{tabular}{l||ccc|r}
\toprule
\colhead{Model} & \colhead{$\log p$} & \colhead{$\feh_0$}& \colhead{$\log\text{M}$} & \colhead{$\Delta$AIC} \\
\midrule
Leaky Box & $-2.53^{+0.05}_{-0.05}$ & \dots & \dots & $+$5.1 \\
Pre Enriched & $-2.56^{+0.05}_{-0.05}$ & $-4.46^{+0.31}_{-0.34}$ & \dots & $+$4.8 \\
Extra Gas & $-2.54^{+0.04}_{-0.05}$ & \dots & $0.05^{+0.08}_{-0.04}$ & $+$7.3 \\
\bottomrule
\end{tabular}
\tablecomments{Lower $\Delta$AIC means it is preferred over the Gaussian model. All models fit worse than a Gaussian model.}
\end{table}

\begin{figure*}
    \centering
    \includegraphics[width=\linewidth]{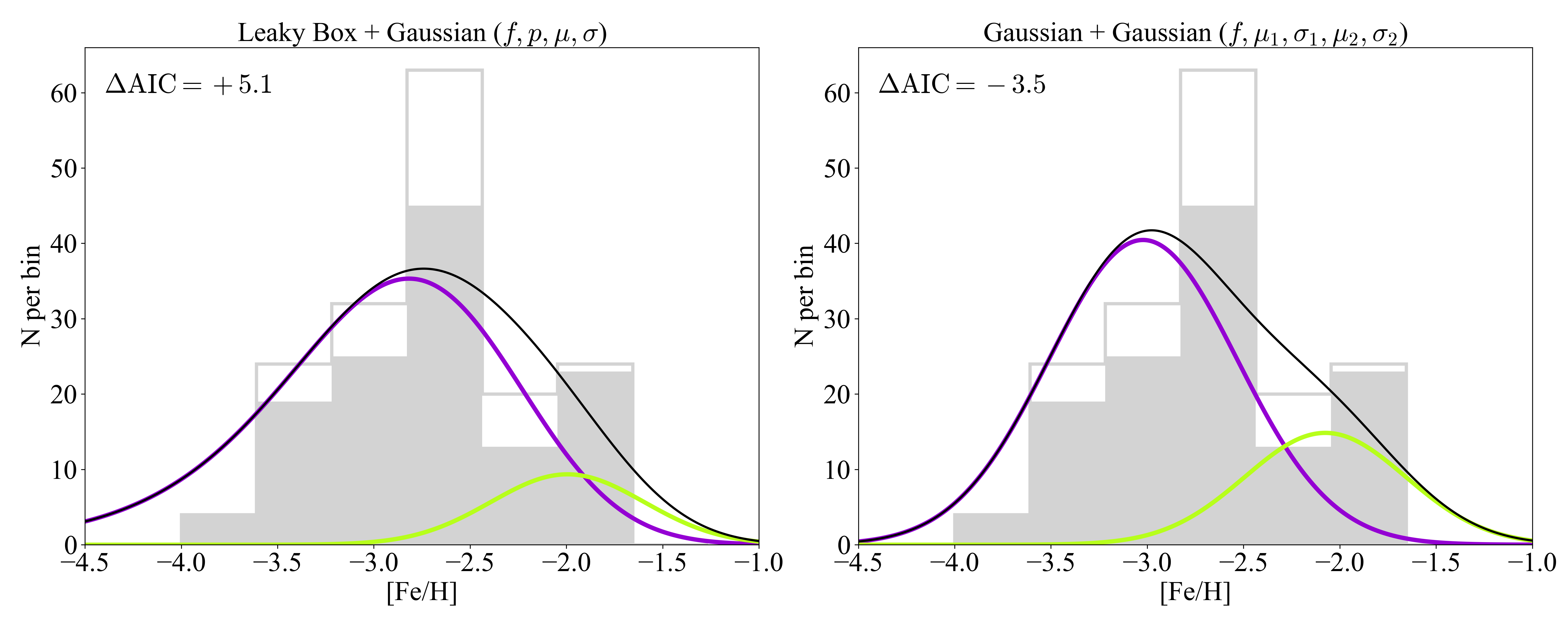}
    \caption{We fit two-component mixture models to test the bimodality of the MDF. Left: The Leaky Box (purple) + Gaussian (green) mixture model. Right: The Gaussian (purple) + Gaussian (green) mixture model. The black line is the result of Equation \ref{pdf_eq}. The $\Delta$AIC values are compared to the single Gaussian MDF. The Gaussian + Gaussian MDF is preferred over a single Gaussian MDF. The fits are convolved with the typical uncertainty of $\sim 0.4$ dex. The metal-poor peak in purple is shifted to the left due to the presence of upper limits.}
    \label{fig:complete_histogram}
\end{figure*}

\begin{table*}
\centering
\caption{\label{tab:bimodal_params} Bimodal Model Parameters}
\begin{tabular}{rc||cccccc|r}
\toprule
\colhead{Model} & \colhead{Uncertainty} & \colhead{$f$} & \colhead{$\log p$}& \colhead{$\mu_1$} & \colhead{$\sigma_1$} & \colhead{$\mu_2$} & \colhead{$\sigma_2$} & \colhead{$\Delta$AIC} \\
\midrule
Leaky Box + Gaussian  &  $e_\text{total}$ & $0.86_{-0.07}^{+0.07}$ & $-2.72_{-0.10}^{+0.11}$ & $-1.99_{-0.11}^{+0.10}$ & $0.23^{*}$ & \nodata & \nodata & $+$5.1 \\
Gaussian + Gaussian &  $e_\text{total}$ & 0.76$_{-0.08}^{+0.07}$ & \nodata & $-3.02_{-0.08}^{+0.08}$ & $0.29_{-0.06}^{+0.08}$ & $-2.08_{-0.09}^{+0.09}$ & $0.25^{*}$ & $-$3.5 \\
\hline
Leaky Box + Gaussian &  $e_\text{spec}$ & 0.86$_{-0.04}^{+0.03}$ & $-2.73_{-0.06}^{+0.06}$ & $-1.91_{-0.04}^{+0.04}$ & $0.10_{-0.03}^{+0.04}$ & \nodata & \nodata & $-$2.6 \\
Gaussian + Gaussian &  $e_\text{spec}$ & 0.85$_{-0.04}^{+0.04}$ & \nodata & $-2.90_{-0.05}^{+0.05}$ & $0.38_{-0.04}^{+0.06}$ & $-1.93_{-0.06}^{+0.05}$ & $0.21^{*}$ & $-$8.1 \\
\bottomrule
\end{tabular}
\tablecomments{AIC is only comparable with the same data. $\Delta$AIC are relative to the single Gaussian model. A lower $\Delta$AIC indicates a better fit. The $^*$ indicates a 90th percentile upper limit. \\ }
\end{table*}

Fitting analytical models to the MDF allows us to learn about the gas flows in the system  \citep{Kirby+2011}. The Gaussian distribution parameters have no physical interpretation whereas chemical evolution (CE) model parameters do. Following \cite{Kirby+2011}, the different models we fit to the MDF are Leaky Box, Pre-Enriched, and Extra Gas models. The simplest model is the Leaky Box model, parameterized by the effective yield, $p$. It describes a system that starts with a finite amount of pristine gas with $\feh_0=-\infty$, allows gas to leave, and does not accrete any other external gas \citep{Pagel1997}. 
A high effective yield could point to a system that retained more of its metals due to a deeper potential well or was more efficient at turning gas into stars.
The Pre-Enriched model is the Leaky Box with the addition of an initial non-zero metallicity to the gas it starts with,$\feh_0$.
The Extra Gas model, also known as the best accretion model in \cite{Lynden-Bell1975}, is a system that both allows the gas to leave and accretes external metal-free gas. The amount of gas is described in the model with the parameter, \textit{M}. The Extra Gas model becomes the Leaky Box model for \textit{M}=1. Table \ref{tab:gce_params} shows the best-fit parameters and the $\Delta$AIC value, which is compared to the AIC value of a single Gaussian MDF (AIC $=279.4$). The CE models are not preferred over a single Gaussian model. Figure \ref{fig:gce_gauss_plot} shows the CE models compared to the Gaussian MDF. 
\begin{figure*}
    \centering
    \includegraphics[width=\linewidth]{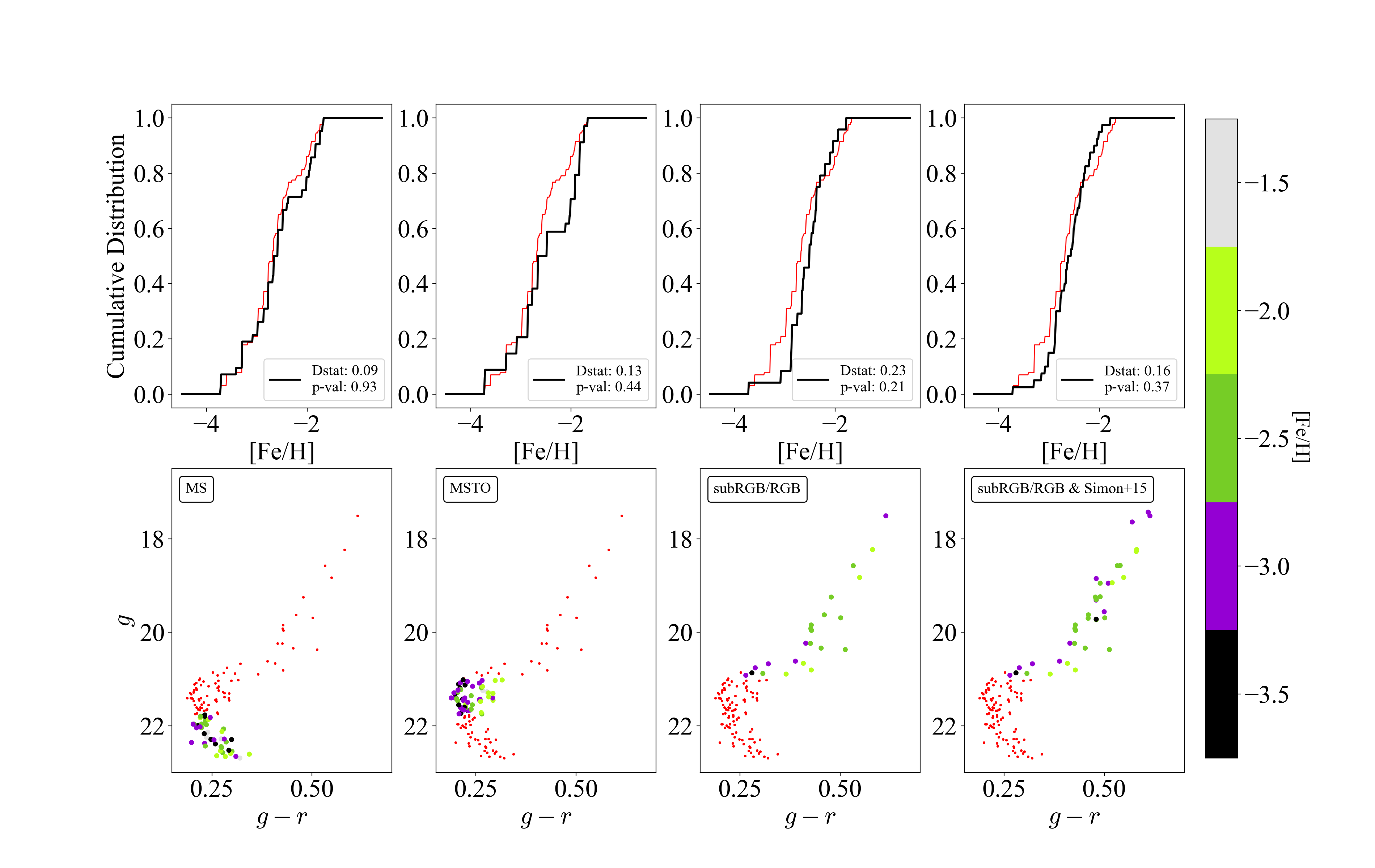}
    \caption{Top: The CDF of the entire sample (red line) compared to the CDF along the MS, MSTO, sub-RGB/RGB regions (black lines). We performed a KS test to determine if there was a bimodality in each region. We report the D-statistic and p-value that demonstrate the distributions in each region are comparable to the overall distribution. The rightmost CDF and CMD include RGB stars from \cite{Simon+2015}. Bottom: The CMD of the different regions colored by metallicity. The entire sample is plotted in small red circles.}
    \label{fig:ks_test}
\end{figure*}

\begin{figure*}
    \centering
    \includegraphics[width=\linewidth]{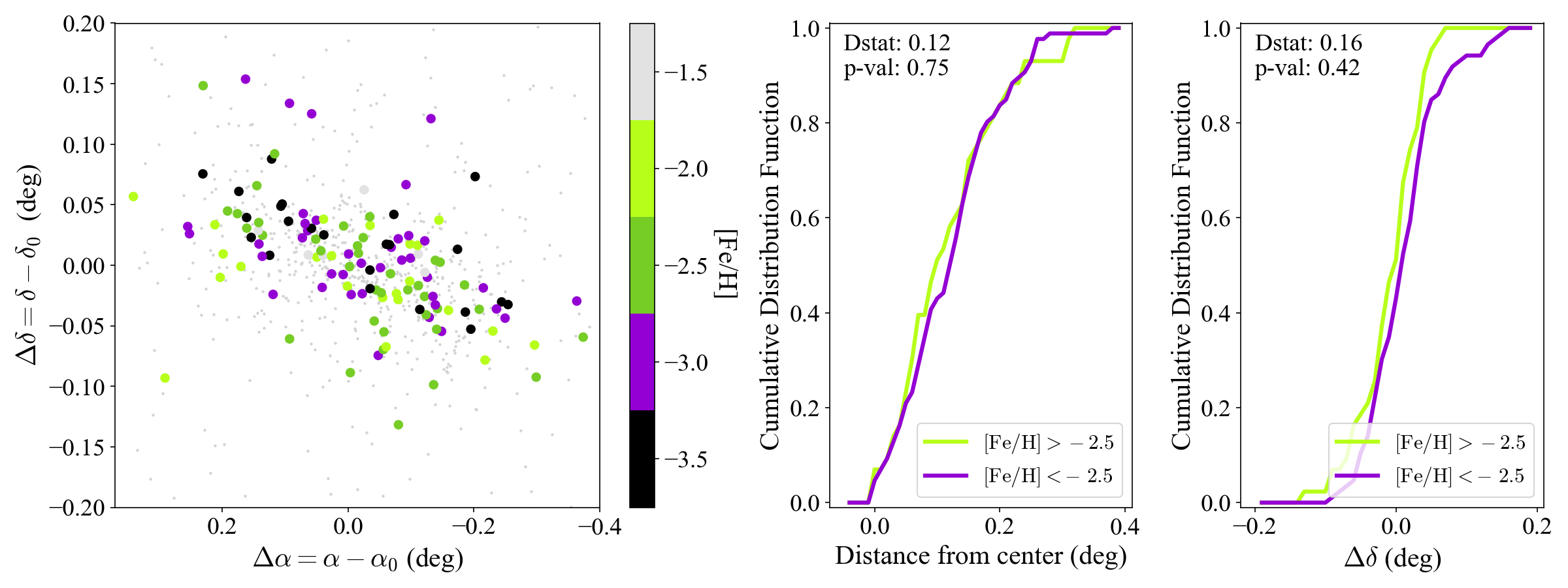}
    \caption{Left: The spatial distribution of stars in Reticulum II colored by metallicity. All stars in the photometric catalogs are plotted in small gray circles.  Center: The cumulative distribution function of the metal-poor stars ( [Fe/H] $< -2.5$) and metal-rich stars ([Fe/H] $> -2.5$) as a function of radial distance. A KS test is performed on the two distributions, with a large p-value indicating there is no spatial separation between the metal-poor and metal-rich populations. Right: The cumulative distribution function of the metal-poor stars ( [Fe/H] $< -2.5$) and metal-rich stars ([Fe/H] $> -2.5$) as a function of $\Delta \delta$. The result of the KS test shows that there is no significantly preferred direction along the $\Delta \delta$ direction for the metal-poor and metal-rich stars.}
    \label{fig:RADEC_CDF}
\end{figure*}

\subsection{Bimodal Model Fitting}
\label{testing_bimodal}
Visually, the MDF shows a potentially bimodal distribution. We fit two-component mixture models (MMs) to test the significance of the bimodality. When we fit all CE models, the $\Delta$AIC values are separated by $\lesssim 2$, suggesting they all fit equally well. 
Therefore, we continue with combinations of the simplest models for the bimodal distributions.
We fit a Leaky Box + Gaussian MM and a Gaussian + Gaussian MM to the MDF.  For these models, we define the mixing fraction, $f$, as the fraction in the lower metallicity peak. We set an extra prior that makes the two component peaks at least 0.4 dex apart. The probability distributions of the MMs follow:
\begin{equation}
    \label{pdf_eq}
    P_M = \\ f \cdot P_\text{1}(z\mid\theta) + (1-f) \cdot P_\text{2}(z\mid \theta),
\end{equation}

where $P_1(z\mid\theta)$ and $P_2(z\mid\theta)$ are the probability distribution functions of the lower and higher metallicity peaks, respectively.

The uncertainties in the \cite{Beers+1999} calibration are not well characterized, affecting $75\%$ of our sample that are dominated by systematic uncertainties. It is unclear whether the sources of the uncertainty are from a zero-point calibration, stellar parameter trends, or intrinsic variability of the metallicities. An updated calibration could potentially reduce the systematic uncertainty. 
To explore a best-case scenario, we run each model separately with the total metallicity uncertainty ($e_\text{total} \sim$ 0.4) and with only the spectroscopic metallicity uncertainty ($e_\text{spec} \sim$ 0.2). Appendix \ref{appendix} shows the posterior distributions of the model parameters and Table \ref{tab:bimodal_params} shows the best-fit parameter values for each mixture model and their $\Delta$AIC values. The $\Delta$AIC values are relative to the Gaussian MDF (AIC $=279.4$). Figure \ref{fig:complete_histogram} shows the mixture model fits convolved with the typical uncertainty of $\sim 0.4$ dex. The $\Delta$AIC values indicate a Leaky Box + Gaussian MM is only preferred over a single Gaussian model when using the spectroscopic uncertainties. The Gaussian + Gaussian MM is preferred in both uncertainty cases. It is slightly preferred with the total uncertainty, but more strongly preferred with the spectroscopic uncertainty, supporting the presence of bimodality in the MDF.

We perform a Kolmogorov–Smirnov (KS) test on the MS, MSTO, and sub-RGB/RGB regions to test our confidence in the bimodality by comparing the cumulative metallicity distributions of our entire sample with those of the different regions. Shown in Figure \ref{fig:ks_test}, the p-values suggest that the entire sample and the MS, MSTO, and sub-RGB/RGB regions are each derived from the same parent distribution. The rightmost plots show the CDF and CMD including stars from \cite{Simon+2015}, increasing our confidence in the bimodality of the sub-RGB/RGB region from the high p-value. 

Mergers could affect the spatial separation of metal-poor and metal-rich populations, as possibly seen in the Sculptor dwarf galaxy \citep{Benitez-Llambay2016, Arroyo-Polonio2024, Barbosa2025}.
In Figure \ref{fig:RADEC_CDF}, we show the spatial distribution of our sample colored by metallicity. We plot the CDFs of metal-poor stars ([Fe/H] $< -2.5$) and of metal-rich stars ([Fe/H] $> -2.5$). Visually, it seems like the most metal-poor stars could be biased to the northern half of Ret~II. However, we find no significant spatial separation in the metal-poor and metal-rich populations, along either the radial or the $\Delta \delta$ directions.
\section{DISCUSSION}
\label{discussion}

\subsection{Interpreting the Bimodality of the Ret II MDF}
We report the first bimodal MDF seen in any UFD \footnote{The Segue 1 UFD was reported to have a wide dispersion or multi-modal metallicity distribution from the range of 7 RGB metallicities spanning more than 2 dex \citep{Frebel2014, Webster2016}. Bissonette et al. (in prep.) has recently confirmed that Segue 1 has a Gaussian MDF with $N=49$.}.
There is a peak at $\langle\feh_1\rangle = -3.02 \pm 0.08$ with $ 76\%$ of the stars and another peak at 
$\langle\feh_2\rangle = -2.08 \pm 0.09$ with $ 24 \%$ of the stars. 
The MDF has a metallicity gap of $\Delta\text{[Fe/H]}_\text{gap} = 0.94$ dex. 
\cite{Ting&Ji2024} proposed a method to predict the observed metallicity gap in an MDF, due to quiescent periods between star formation events, for a range of dwarf galaxy masses. The stars from the first burst undergo Type Ia supernovae and eject iron, enriching the gas and resulting in the subsequent burst of stars forming with a higher metallicity.
Using the SFH in \cite{Simon+2023}, they determined a $\Delta \feh_\text{gap}= $ 1.09 dex for Ret II, consistent with the gap in our observed MDF, $\Delta \feh_\text{gap} \sim 1.0$ dex. This supports their theory of predicting metallicity gaps in dwarf galaxies.

The current SFH for Reticulum II in \cite{Simon+2023} prefers a two-burst star formation model. The best-fit instantaneous burst model has a burst at 14.1 Gyr containing 87.5\% of stars and a second burst 3.4 Gyr later forming the final 12.5\% of stars. They constrain SFH models by selecting isochrones that follow the single-mode MDF in \cite{Simon+2015}. The percentage of stars in each burst from the SFH is consistent with the results of our bimodal MDF analysis, where we have a mixing fraction of $f= 0.76_{-0.08}^{+0.07}$ when using total uncertainties, and with only spectroscopic uncertainties we have $f= 0.85_{-0.04}^{+0.04}$. Further constraining the models going into the SFH to follow a bimodal MDF would result in a more accurate distribution of isochrones, and potentially the SFH. 

\begin{figure}
    \centering
    \includegraphics[width=\linewidth]{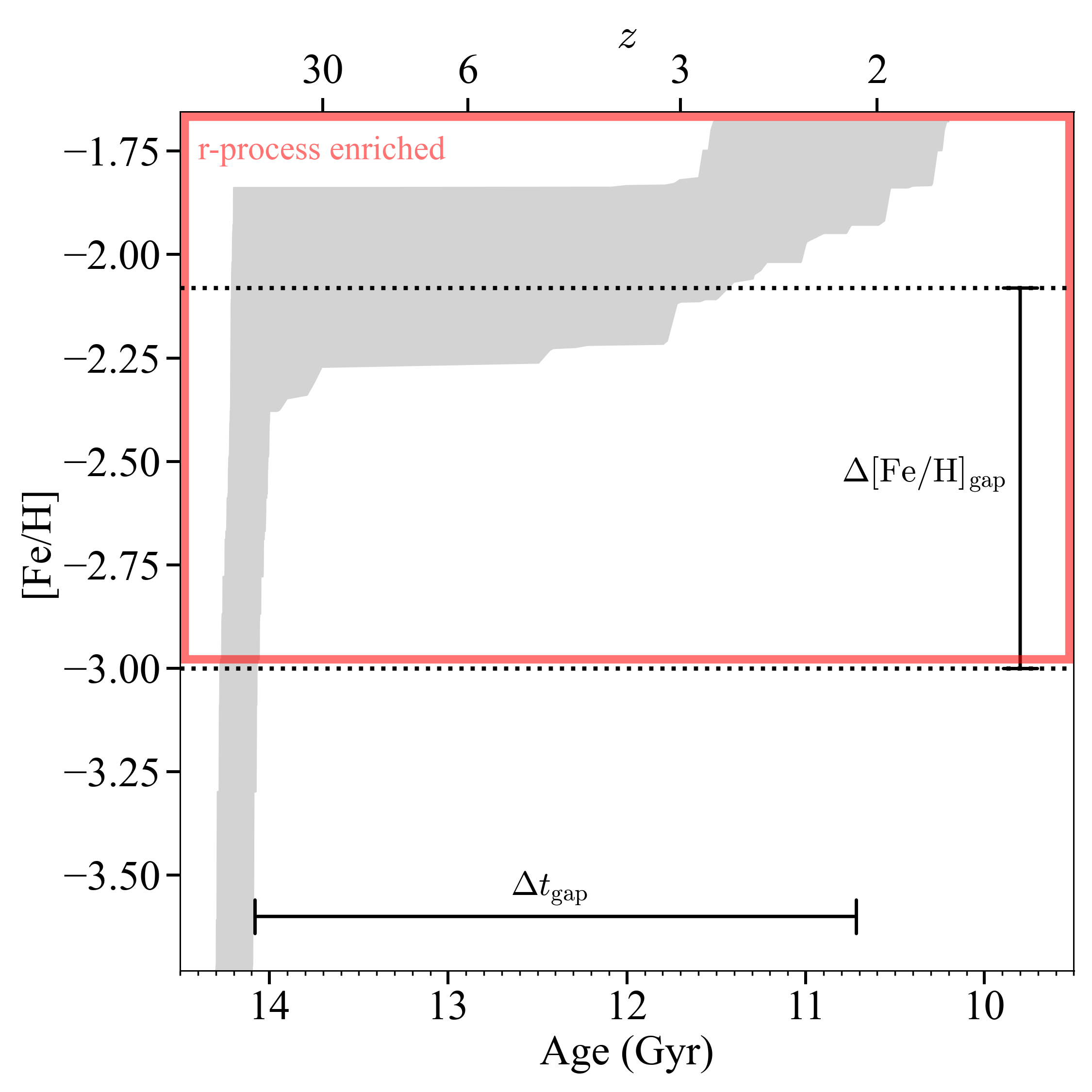}
    \caption{The age-metallicity relation using the cumulative SFH in \cite{Simon+2023} interpolated with 0.1 steps to the CDF of our metallicity distribution. The red box shows the 72\% of stars that are r-process enriched from \cite{Ji+2023}. There is some star formation before the r-process event enriches the gas in the system. The second burst that contains $\sim 20\%$ of stars should all be r-process enriched. A $\Delta\text{[Fe/H]}_\text{gap} \sim 1.0$ dex is related to $\Delta t_\text{gap} \sim 3$ Gyr. Note that the ages are computed on a scale where the age of M92 is 13.2 Gyr.}  
    \label{fig:feh_age}
\end{figure}

As an initial estimate, we determined the age-metallicity relation (AZR) using the cumulative SFH in \cite{Simon+2023} interpolated with 0.1 steps to the CDF of our bimodal metallicity distribution, shown in Figure \ref{fig:feh_age}. The AZR combines our MDF with the SFH by assuming that metallicity increases monotonically with time, with the metal-rich peak consisting of younger stars. The AZR suggests that the metallicity gap of 1 dex is related to an age gap of 3 Gyr, and therefore that there is star formation after reionization.

\begin{figure*}
    \centering 
    \includegraphics[width=\linewidth]{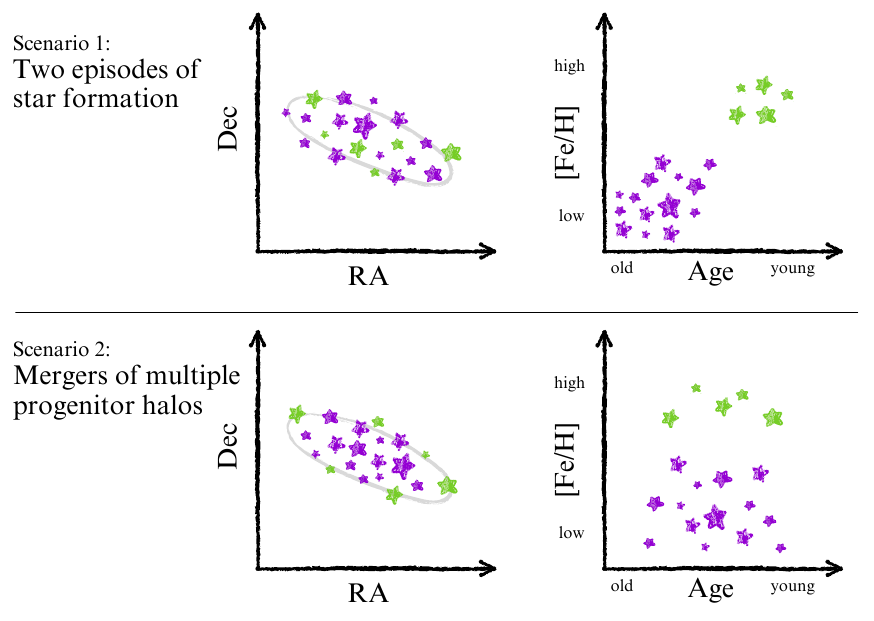}
    \caption{Observables from the two possible scenarios that result in a bimodal MDF. Scenario 1, which we prefer, has a mixed spatial distribution and a correlated AZR. Scenario 2 has the accreted metal-rich stars on the outer parts of the galaxy and no correlation in the AZR. }
    \label{fig:bimodal_scenario}
\end{figure*}

Other independent analyses of the formation history of Reticulum II indicate similar results. \cite{Alexander+2023} predicted a bimodal metallicity distribution of Ret II from modeling inhomogeneous chemical evolution. They interpret the higher metallicity mode as a consequence of implementing a delay-time distribution for Type Ia SNe in their modeling. \cite{Ji+2023} also found independent evidence of bursty star formation in Reticulum II from well-mixed Barium abundances in r-process enriched stars. 

Although we prefer the two-burst star formation scenario, the accretion of multiple progenitor halos could also result in a bimodal MDF. Simulations suggest that UFDs may form their stars in multiple smaller halos from different environments that eventually merge \citep{Simpson2013, Jeon2017} and can be dominated by late-time dry mergers \citep{Rey2019, Andersson2025}. The progenitors evolve to have distinct MDFs that are imprinted on the overall MDF once accreted \citep{Ko2024}. For example, a halo at $z=0$ can have accreted many progenitor halos with varying metallicities. This is a possible interpretation for the bimodal MDF. Although \cite{Ko2024} did not find a bimodality in any resulting halo MDF, we note that it could be possible if there were not many accretions that could wash out the bimodality in the final MDF. The accreted population is more likely to be more metal-poor, however, to agree with the Ret II MDF, the accreted population comprising of the minority of stars is more metal-rich. This scenario could be distinguishable from the two-burst star formation scenario from the AZR and the spatial distribution. Figure \ref{fig:bimodal_scenario} shows the expected results of the spatial metallicity distribution and AZR for the two scenarios. We investigated the spatial distribution of our sample in Figure \ref{fig:RADEC_CDF} and found no spatial separation between the metal-rich and metal-poor populations suggesting a preference for the two-burst scenario, but a full AZR with independent measurements of age and metallicity would be decisive.

Another possible scenario that could affect the MDF is related to gas accretion of the intergalactic medium (IGM). \cite{Ahvazi2024} introduced an IGM metallicity model in their simulations, reproducing the observed mass-metallicity relation (MZR) at small scales. However, \cite{Wheeler2025} found that the amount of IGM accretion is not sufficient to affect the MDF. While more work is needed to study UFD MDFs, we expect the AZR to be distinguishable from the other scenarios.  
\subsection{r-process in Ret II}
Ret II was discovered to have experienced a rare and prolific r-process event that produced large amounts of r-process elements such as Europium and Barium \citep{Ji+2016a, Ji+2016c, Roederer2016}. There is a current debate about the possible sites for r-process nucleosynthesis \citep{Frebel&Ji2023}. The two most likely astrophysical sites are neutron star mergers (NSM) and rare core-collapse supernovae (CCSNe), which occur at drastically different delay times, with the former taking up to billions of years and the latter all occurring after a few million years. \cite{Simon+2023} constrained the timing of the nucleosynthetic event by studying the extended star formation history using that it must have occurred after 28$\%$ of non-r-process-enhanced stars had formed \citep{Ji+2023} and found that the event occurred within 500 $\pm$ 200 Myr of the formation of the first stars in Ret II. 
The AZR in Figure \ref{fig:feh_age} shows that the second burst of stars must all be r-process enriched. With the second burst of stars having higher [Fe/H], we should see a decrease in [r/Fe] when there is an increase in Fe. A possible indication of this trend in the literature is that the star DES J033548$-$540349 was relatively metal-rich with $\feh = -2.19$ but somewhat less enhanced in neutron-capture abundances than the metal-poor stars in \cite{Ji+2016c}.

\subsection{Implications of the Formation History and Evolution of Ret II and other UFDs}
\label{formation_hist}
Star formation on the smallest scales is quenched or suppressed during the epoch of reionization \citep{Bullock2000,Brown2014,Weisz2014,Jeon2017,RodriguezWimberly2019}, however, it is unclear if this effect is homogeneous for all UFDs. 
The SFHs of Magellanic Cloud (MC) satellites are suggested to continue 600 Myr longer than the Milky Way (MW) satellites, due to a weaker local ionization field during reionization, emphasizing the importance of environmental effects on star formation in UFDs \citep{Sacchi+2021, Kim2023, Durbin2025}. 
\cite{Savino2023} shows that compared to MW satellites, M31 satellites continue to form stars longer after reionization, concluding that M31 UFDs are not as affected by reionization. They note that the M31 UFDs in their study are more massive than the MW satellites, which could affect the amount of gas retained. A study of Pegasus W, an isolated UFD outside the virial radius of M31, also shows the impact of environment on star formation as it formed $\sim 50 \%$ of its stars after $z\sim 6$ \citep{McQuinn2023_PegasusW}. These studies suggest that reionization is not homogeneously quenching star formation, and that environment is also an important factor. In this study, we find that reionization did not completely quench the star formation in Reticulum II, with a stellar mass of $M_* = 10^{3.51\pm0.04} M_\odot$, which simulations predict is below the minimum mass threshold of galaxies massive enough to retain gas and form stars after reionization \citep[e.g.,][]{Kravtsov2004}. The AZR in Figure \ref{fig:feh_age} shows a second burst post-reionization, suggesting the reignition of star formation. 
Therefore, Ret II may have finally self-quenched from internal mechanisms such as stellar or supernova feedback, not by reionization \citep[e.g.,][]{Gallart2021}. With more well-populated MDFs in UFDs, we can detect metallicity gaps related to gaps in age, revealing the importance of reionization versus internal quenching mechanisms.

\section{CONCLUSION}
\label{conclusion}
The star formation histories of ultra-faint satellites of the Milky Way have revealed that most of their stars formed in the early Universe before $z \sim 6$, making UFDs extremely ancient systems \citep{Brown2014, Jenkins2021, Simon+2023}. However, studies suggest that there is star formation after reionization highlighting the importance of the environment in fully quenching these systems \citep{Sacchi+2021,Savino2023, McQuinn2023_PegasusW}. To study the star formation histories of UFDs, we need well-populated MDFs. The MDFs for large dwarf galaxies around the Milky Way have been well studied \citep{Kirby+2011}, however, UFDs are more difficult to get well-populated spectroscopic MDFs and are not as extensively studied. 

In this work, we increased the number of spectroscopic metallicities in the Reticulum II UFD by $\sim$ 6.5 times. This is the most populated MDF of any UFD. We detect a bimodal metallicity distribution with a low metallicity peak at $\feh\sim -3.0$ with about $80\%$ of the stars and a high metallicity peak at $\feh\sim -2.1$ with about $20\%$ of the stars. This is the first bimodal MDF in any UFD. The bimodal MDF is consistent with the current two-burst star formation history in Ret II, which forms $\sim 80 \%$ of its stars in one burst during the early universe and $\sim 20\%$ of its stars a few Gyrs later, after reionization \citep{Simon+2023}. The age-metallicity relation from the two-burst SFH in the literature and our bimodal MDF relates the metallicity gap of $\sim 1.0$ dex to the 3 Gyr age gap, suggesting that there is star formation after reionization. However, we need to get stellar ages using our metallicities to be able to distinguish this scenario from a multiple progenitor halo accretion scenario. Further, with more well-populated UFD MDFs, we will be able to investigate the relative importance between reionization and internal quenching on the smallest scale galaxies.

\section*{Acknowledgements}
AML acknowledges support from the U.S. National Science Foundation (NSF) Graduate Research Fellowship Program.
APJ acknowledges support from the Alfred P. Sloan Research Fellowship and the National Science Foundation grant AST-2307599.
We thank the support astronomers and staff at Las Campanas Observatory (LCO) who made gathering the data possible. This paper made use of NASA’s Astrophysics Data System Bibliographic Services. 

\bibliography{main}{}
\bibliographystyle{aasjournal}

\appendix

\onecolumngrid
\section{Metallicity Table}
\label{metallicity_tab_appendix}
\begin{longtable}{lccccccccc} 
\caption{\label{tab:measurements}Metallicity Values for Members} \\
\toprule
Star ID & R.A.  & Decl. & SNR/px & EW\textcolor{vibrantpurple}{\textbf{*}} & [Fe/H] & $e_\text{calib}$ & $e_\text{spec}$ & $e_\text{phot}$ & $e_\text{total} $\\ & (h:m:s) & (d:m:s) &  & (\AA) & (dex) & (dex) & (dex) & (dex) & (dex)\\
\midrule
\endfirsthead

\multicolumn{3}{c}%
{\tablename\ \thetable\ -- \textit{Continued from previous page}} \\
\toprule
Star ID & R.A.  & Decl. & SNR/px & EW\textcolor{vibrantpurple}{\textbf{*}} & [Fe/H] & $e_\text{calib}$ & $e_\text{spec}$ & $e_\text{phot}$ & $e_\text{total} $\\ & (h:m:s) & (d:m:s) &  & (\AA) & (dex) & (dex) & (dex) & (dex) & (dex)\\
\midrule
\endhead

\midrule \multicolumn{3}{r}{\textit{Continued on next page}} \\
\endfoot

\bottomrule
\endlastfoot

\startdata
 HST\_001 & 03:36:16.96 & $-$53:57:31.75 & 5.2 & 0.45 $\pm$ 0.43 & $-$3.72 $\pm$ 0.56 & 0.30 & 0.48 & 0.00 & 0.56  \\
 HST\_003 & 03:36:15.81 & $-$53:57:16.23 & 11.6 & 1.32 $\pm$ 0.20 & $-$2.59 $\pm$ 0.36 & 0.32 & 0.16 & 0.00 & 0.36  \\
 HST\_007 & 03:35:42.04 & $-$54:01:25.47 & 7.3 & 1.59 $\pm$ 0.30 & $-$2.49 $\pm$ 0.38 & 0.33 & 0.18 & 0.00 & 0.38  \\
 HST\_011 & 03:35:49.49 & $-$54:00:51.02 & 11.4 & 1.39 $\pm$ 0.18 & $-$2.58 $\pm$ 0.37 & 0.34 & 0.13 & 0.00 & 0.36  \\
 HST\_012 & 03:35:39.41 & $-$54:00:49.38 & 15.0 & 3.83 $\pm$ 0.20 & $-$2.21 $\pm$ 0.27 & 0.27 & 0.02 & 0.00 & 0.27  \\
 HST\_015 & 03:35:39.52 & $-$54:00:23.40 & 7.2 & 1.59 $\pm$ 0.33 & $-$2.48 $\pm$ 0.41 & 0.35 & 0.21 & 0.00 & 0.41  \\
 HST\_017 & 03:35:30.35 & $-$54:00:16.56 & 8.1 & 0.60 $\pm$ 0.28 & $-$3.30 $\pm$ 0.48 & 0.30 & 0.37 & 0.00 & 0.48  \\
 HST\_023 & 03:36:00.11 & $-$54:01:30.19 & 3.1 & 1.44 $\pm$ 0.78 & $-$2.67 $\pm$ 0.68 & 0.37 & 0.57 & 0.00 & 0.68  \\
 HST\_026 & 03:35:57.04 & $-$54:01:17.40 & 4.1 & 0.74 $\pm$ 0.60 & $-$3.30 $\pm$ 0.67 & 0.30 & 0.60 & 0.00 & 0.67  \\
 HST\_029 & 03:36:03.20 & $-$54:01:04.88 & 9.2 & 0.67 $\pm$ 0.25 & $-$3.19 $\pm$ 0.43 & 0.29 & 0.32 & 0.00 & 0.43  \\
 HST\_031 & 03:36:01.74 & $-$54:00:57.95 & 10.8 & 0.71 $\pm$ 0.22 & $-$3.30 $\pm$ 0.40 & 0.30 & 0.27 & 0.00 & 0.40  \\
 HST\_033 & 03:36:04.19 & $-$54:00:43.64 & 13.4 & 1.16 $\pm$ 0.16 & $-$2.78 $\pm$ 0.39 & 0.35 & 0.17 & 0.00 & 0.39  \\
 HST\_035 & 03:36:10.51 & $-$54:00:36.78 & 11.1 & 0.65 $\pm$ 0.22 & $-$3.30 $\pm$ 0.42 & 0.30 & 0.29 & 0.00 & 0.41  \\
 HST\_037 & 03:35:59.94 & $-$54:00:33.92 & 6.4 & 1.03 $\pm$ 0.37 & $-$2.78 $\pm$ 0.50 & 0.33 & 0.37 & 0.00 & 0.50  \\
 HST\_039 & 03:35:57.15 & $-$54:00:30.70 & 5.3 & 2.82 $\pm$ 0.60 & $-$1.95 $\pm$ 0.29 & 0.23 & 0.17 & 0.00 & 0.29  \\
 HST\_040 & 03:36:04.90 & $-$54:00:14.10 & 10.3 & 1.08 $\pm$ 0.21 & $-$2.78 $\pm$ 0.39 & 0.33 & 0.20 & 0.00 & 0.38  \\
 HST\_047 & 03:36:24.73 & $-$54:01:24.82 & 17.5 & 0.35 $\pm$ 0.13 & $-$3.61 $\pm$ 0.36 & 0.27 & 0.24 & 0.00 & 0.36  \\
 HST\_049 & 03:36:20.41 & $-$54:01:18.44 & 8.9 & 1.29 $\pm$ 0.25 & $-$2.59 $\pm$ 0.38 & 0.32 & 0.21 & 0.00 & 0.38  \\
 HST\_050 & 03:36:22.04 & $-$54:01:04.13 & 3.6 & 3.57 $\pm$ 0.91 & $-$1.70 $\pm$ 0.34 & 0.21 & 0.26 & 0.00 & 0.34  \\
 HST\_054 & 03:36:21.87 & $-$54:00:40.64 & 19.6 & 2.72 $\pm$ 0.18 & $-$2.38 $\pm$ 0.32 & 0.32 & 0.05 & 0.00 & 0.32  \\
 HST\_057 & 03:36:26.49 & $-$54:00:25.79 & 8.9 & 0.28 $\pm$ 0.27 & $-$3.61 $\pm$ 0.45 & 0.28 & 0.35 & 0.00 & 0.45  \\
 HST\_059 & 03:36:29.99 & $-$54:00:14.22 & 5.9 & 1.97 $\pm$ 0.29 & $-$2.38 $\pm$ 0.35 & 0.33 & 0.12 & 0.00 & 0.35  \\
 HST\_061 & 03:36:33.74 & $-$54:00:06.42 & 10.5 & 1.09 $\pm$ 0.21 & $-$2.69 $\pm$ 0.38 & 0.31 & 0.21 & 0.00 & 0.38  \\
 HST\_065 & 03:36:29.45 & $-$53:59:08.08 & 8.1 & 0.52 $\pm$ 0.30 & $-$3.42 $\pm$ 0.50 & 0.33 & 0.38 & 0.00 & 0.50  \\
 HST\_066 & 03:36:22.58 & $-$53:58:50.97 & 9.4 & 1.01 $\pm$ 0.23 & $-$2.69 $\pm$ 0.40 & 0.31 & 0.25 & 0.00 & 0.40  \\
 HST\_068 & 03:34:49.27 & $-$54:04:36.72 & 8.2 & 0.31 $\pm$ 0.29 & $-$3.72 $\pm$ 0.46 & 0.30 & 0.36 & 0.00 & 0.47  \\
 HST\_073 & 03:34:56.24 & $-$54:03:54.84 & 3.9 & 0.98 $\pm$ 0.62 & $-$2.99 $\pm$ 0.65 & 0.30 & 0.57 & 0.00 & 0.65  \\
 HST\_076 & 03:35:03.42 & $-$54:03:46.31 & 18.0 & 2.56 $\pm$ 0.18 & $-$2.38 $\pm$ 0.32 & 0.32 & 0.04 & 0.06 & 0.33  \\
 HST\_083 & 03:35:06.07 & $-$54:02:00.25 & 7.4 & 0.70 $\pm$ 0.34 & $-$3.30 $\pm$ 0.50 & 0.30 & 0.40 & 0.00 & 0.50  \\
 HST\_091 & 03:35:28.74 & $-$54:04:29.69 & 4.1 & 3.78 $\pm$ 0.77 & $-$1.84 $\pm$ 0.29 & 0.22 & 0.18 & 0.00 & 0.29  \\
 HST\_093 & 03:35:18.66 & $-$54:04:20.63 & 5.9 & 1.34 $\pm$ 0.36 & $-$2.67 $\pm$ 0.46 & 0.37 & 0.28 & 0.00 & 0.46  \\
 HST\_094 & 03:35:29.38 & $-$54:04:12.10 & 4.2 & 4.00 $\pm$ 0.77 & $-$1.84 $\pm$ 0.28 & 0.22 & 0.18 & 0.00 & 0.28  \\
 HST\_101 & 03:35:25.00 & $-$54:04:00.75 & 9.1 & 1.35 $\pm$ 0.25 & $-$2.50 $\pm$ 0.36 & 0.30 & 0.20 & 0.00 & 0.36  \\
 HST\_105 & 03:35:20.98 & $-$54:03:48.10 & 18.8 & 3.24 $\pm$ 0.18 & $-$2.52 $\pm$ 0.30 & 0.29 & 0.05 & 0.00 & 0.29  \\
 HST\_106 & 03:35:24.32 & $-$54:03:35.25 & 9.2 & 3.00 $\pm$ 0.36 & $-$2.02 $\pm$ 0.26 & 0.25 & 0.09 & 0.00 & 0.26  \\
 HST\_109 & 03:35:17.56 & $-$54:03:23.44 & 10.4 & 0.90 $\pm$ 0.23 & $-$2.99 $\pm$ 0.40 & 0.30 & 0.25 & 0.00 & 0.39  \\
 HST\_113 & 03:35:18.52 & $-$54:03:08.69 & 3.7 & 4.91 $\pm$ 0.98 & $-$1.75 $\pm$ 0.31 & 0.21 & 0.23 & 0.00 & 0.31  \\
 HST\_117 & 03:35:12.71 & $-$54:02:38.65 & 6.4 & 1.36 $\pm$ 0.34 & $-$2.59 $\pm$ 0.42 & 0.32 & 0.26 & 0.00 & 0.41  \\
 HST\_120 & 03:35:14.54 & $-$54:02:33.11 & 15.0 & 3.67 $\pm$ 0.27 & $-$2.31 $\pm$ 0.28 & 0.28 & 0.04 & 0.00 & 0.28  \\
 HST\_121 & 03:35:27.33 & $-$54:02:32.28 & 5.4 & 0.79 $\pm$ 0.42 & $-$2.98 $\pm$ 0.57 & 0.32 & 0.47 & 0.00 & 0.57  \\
 HST\_125 & 03:35:24.02 & $-$54:02:26.62 & 30.4 & 1.28 $\pm$ 0.08 & $-$2.88 $\pm$ 0.39 & 0.38 & 0.05 & 0.00 & 0.38  \\
 HST\_128 & 03:35:21.03 & $-$54:02:14.16 & 11.4 & 4.33 $\pm$ 0.29 & $-$1.68 $\pm$ 0.22 & 0.20 & 0.07 & 0.00 & 0.21  \\
 HST\_132 & 03:35:21.34 & $-$54:01:48.41 & 5.9 & 2.66 $\pm$ 0.50 & $-$2.11 $\pm$ 0.30 & 0.26 & 0.14 & 0.00 & 0.29  \\
 HST\_133 & 03:35:24.20 & $-$54:01:44.67 & 9.9 & 2.00 $\pm$ 0.21 & $-$2.23 $\pm$ 0.28 & 0.26 & 0.10 & 0.00 & 0.28  \\
 HST\_135 & 03:35:18.56 & $-$54:01:35.15 & 10.9 & 1.16 $\pm$ 0.22 & $-$2.78 $\pm$ 0.41 & 0.35 & 0.21 & 0.00 & 0.41  \\
 HST\_136 & 03:35:28.67 & $-$54:01:29.06 & 11.7 & 0.81 $\pm$ 0.18 & $-$3.09 $\pm$ 0.40 & 0.35 & 0.19 & 0.00 & 0.40  \\
 HST\_139 & 03:35:34.61 & $-$54:04:23.66 & 14.4 & 3.47 $\pm$ 0.23 & $-$2.11 $\pm$ 0.26 & 0.26 & 0.03 & 0.00 & 0.26  \\
 HST\_140 & 03:35:42.42 & $-$54:04:12.34 & 6.5 & 1.17 $\pm$ 0.34 & $-$2.78 $\pm$ 0.45 & 0.33 & 0.30 & 0.00 & 0.44  \\
 HST\_141 & 03:35:35.16 & $-$54:04:09.61 & 12.4 & 1.52 $\pm$ 0.19 & $-$2.49 $\pm$ 0.37 & 0.34 & 0.14 & 0.00 & 0.37  \\
 HST\_143 & 03:35:37.08 & $-$54:04:01.21 & 63.5 & 3.19 $\pm$ 0.05 & $-$2.65 $\pm$ 0.28 & 0.27 & 0.00 & 0.07 & 0.28  \\
 HST\_145 & 03:35:39.42 & $-$54:03:57.27 & 15.9 & 0.65 $\pm$ 0.15 & $-$3.30 $\pm$ 0.38 & 0.32 & 0.21 & 0.00 & 0.38  \\
 HST\_164 & 03:35:49.66 & $-$54:03:15.44 & 8.8 & 1.04 $\pm$ 0.27 & $-$2.78 $\pm$ 0.42 & 0.33 & 0.26 & 0.00 & 0.42  \\
 HST\_165 & 03:35:31.61 & $-$54:03:12.93 & 8.4 & 1.43 $\pm$ 0.26 & $-$2.67 $\pm$ 0.42 & 0.37 & 0.20 & 0.00 & 0.42  \\
 HST\_167 & 03:35:39.47 & $-$54:03:01.96 & 13.9 & 0.62 $\pm$ 0.17 & $-$3.30 $\pm$ 0.39 & 0.30 & 0.24 & 0.00 & 0.39  \\
 HST\_169 & 03:35:35.45 & $-$54:02:54.85 & 17.4 & 1.24 $\pm$ 0.13 & $-$2.87 $\pm$ 0.41 & 0.40 & 0.08 & 0.00 & 0.41  \\
 HST\_170 & 03:35:47.23 & $-$54:02:51.72 & 5.2 & 1.32 $\pm$ 0.41 & $-$2.67 $\pm$ 0.49 & 0.37 & 0.33 & 0.00 & 0.49  \\
 HST\_176 & 03:35:43.84 & $-$54:02:12.02 & 7.6 & 1.47 $\pm$ 0.28 & $-$2.58 $\pm$ 0.39 & 0.34 & 0.19 & 0.00 & 0.39  \\
 HST\_178 & 03:35:31.36 & $-$54:01:54.28 & 15.1 & 0.93 $\pm$ 0.16 & $-$3.09 $\pm$ 0.39 & 0.35 & 0.16 & 0.00 & 0.38  \\
 HST\_179 & 03:35:44.19 & $-$54:01:49.96 & 16.0 & 4.26 $\pm$ 0.19 & $-$2.44 $\pm$ 0.27 & 0.26 & 0.03 & 0.00 & 0.26  \\
 HST\_180 & 03:35:33.21 & $-$54:01:45.23 & 4.2 & 0.58 $\pm$ 0.59 & $-$3.30 $\pm$ 0.65 & 0.33 & 0.56 & 0.00 & 0.65  \\
 HST\_189 & 03:35:57.67 & $-$54:03:53.64 & 6.6 & 0.99 $\pm$ 0.35 & $-$3.09 $\pm$ 0.50 & 0.34 & 0.37 & 0.00 & 0.50  \\
 HST\_194 & 03:35:54.08 & $-$54:03:13.46 & 9.7 & 1.06 $\pm$ 0.24 & $-$2.78 $\pm$ 0.43 & 0.35 & 0.25 & 0.00 & 0.43  \\
 HST\_203 & 03:35:59.69 & $-$54:02:23.50 & 3.3 & 2.54 $\pm$ 1.05 & $-$2.12 $\pm$ 0.51 & 0.26 & 0.44 & 0.00 & 0.51  \\
 HST\_204 & 03:35:57.43 & $-$54:02:18.43 & 4.5 & 4.05 $\pm$ 0.68 & $-$1.68 $\pm$ 0.27 & 0.20 & 0.18 & 0.00 & 0.27  \\
 HST\_205 & 03:36:02.99 & $-$54:02:17.06 & 3.7 & 4.49 $\pm$ 0.92 & $-$1.75 $\pm$ 0.32 & 0.21 & 0.24 & 0.00 & 0.32  \\
 HST\_207 & 03:35:58.16 & $-$54:02:04.73 & 38.8 & 3.15 $\pm$ 0.08 & $-$2.52 $\pm$ 0.30 & 0.29 & 0.03 & 0.00 & 0.29  \\
 HST\_213 & 03:36:16.40 & $-$54:04:14.40 & 4.4 & 1.24 $\pm$ 0.55 & $-$2.87 $\pm$ 0.60 & 0.39 & 0.46 & 0.00 & 0.60  \\
 HST\_218 & 03:36:28.67 & $-$54:02:51.40 & 8.7 & 2.69 $\pm$ 0.38 & $-$2.04 $\pm$ 0.27 & 0.24 & 0.11 & 0.00 & 0.27  \\
 HST\_219 & 03:36:17.69 & $-$54:02:18.07 & 7.4 & 0.61 $\pm$ 0.32 & $-$3.30 $\pm$ 0.51 & 0.30 & 0.41 & 0.00 & 0.51  \\
 HST\_220 & 03:36:35.42 & $-$54:02:14.10 & 6.2 & 2.56 $\pm$ 0.49 & $-$1.97 $\pm$ 0.29 & 0.24 & 0.17 & 0.00 & 0.29  \\
 HST\_222 & 03:36:21.78 & $-$54:01:44.55 & 12.4 & 0.98 $\pm$ 0.18 & $-$2.98 $\pm$ 0.37 & 0.32 & 0.19 & 0.00 & 0.37  \\
 HST\_224 & 03:34:55.58 & $-$54:07:29.68 & 4.1 & 2.81 $\pm$ 0.65 & $-$2.02 $\pm$ 0.31 & 0.25 & 0.20 & 0.00 & 0.32  \\
 HST\_236 & 03:34:52.57 & $-$54:06:03.40 & 9.2 & 3.47 $\pm$ 0.35 & $-$1.93 $\pm$ 0.24 & 0.23 & 0.06 & 0.00 & 0.24  \\
 HST\_239 & 03:35:01.00 & $-$54:05:57.91 & 7.2 & 0.43 $\pm$ 0.35 & $-$3.61 $\pm$ 0.50 & 0.27 & 0.42 & 0.00 & 0.50  \\
 HST\_241 & 03:35:03.09 & $-$54:05:06.90 & 8.3 & 0.64 $\pm$ 0.28 & $-$3.30 $\pm$ 0.48 & 0.32 & 0.36 & 0.00 & 0.48  \\
 HST\_242 & 03:34:57.79 & $-$54:04:58.72 & 7.1 & 1.40 $\pm$ 0.30 & $-$2.58 $\pm$ 0.41 & 0.34 & 0.22 & 0.00 & 0.41  \\
 HST\_243 & 03:34:51.22 & $-$54:04:57.71 & 4.0 & 1.11 $\pm$ 0.58 & $-$2.87 $\pm$ 0.63 & 0.37 & 0.51 & 0.00 & 0.63  \\
 HST\_256 & 03:35:12.16 & $-$54:06:04.30 & 13.8 & 0.98 $\pm$ 0.17 & $-$2.98 $\pm$ 0.37 & 0.32 & 0.17 & 0.00 & 0.36  \\
 HST\_257 & 03:35:14.03 & $-$54:05:58.14 & 26.5 & 1.69 $\pm$ 0.09 & $-$2.67 $\pm$ 0.38 & 0.38 & 0.05 & 0.00 & 0.38  \\
 HST\_260 & 03:35:16.88 & $-$54:05:22.52 & 17.0 & 1.50 $\pm$ 0.14 & $-$2.76 $\pm$ 0.41 & 0.40 & 0.08 & 0.00 & 0.41  \\
 HST\_262 & 03:35:18.02 & $-$54:05:15.12 & 6.6 & 1.00 $\pm$ 0.29 & $-$2.69 $\pm$ 0.43 & 0.31 & 0.30 & 0.00 & 0.43  \\
 HST\_264 & 03:35:09.51 & $-$54:05:01.81 & 6.9 & 3.49 $\pm$ 0.44 & $-$1.93 $\pm$ 0.24 & 0.23 & 0.08 & 0.00 & 0.24  \\
 HST\_266 & 03:35:20.46 & $-$54:04:59.23 & 6.4 & 0.55 $\pm$ 0.37 & $-$3.30 $\pm$ 0.54 & 0.30 & 0.45 & 0.00 & 0.54  \\
 HST\_267 & 03:35:13.74 & $-$54:04:56.67 & 30.1 & 2.90 $\pm$ 0.11 & $-$2.40 $\pm$ 0.30 & 0.30 & 0.00 & 0.00 & 0.30  \\
 HST\_269 & 03:35:46.95 & $-$54:08:07.15 & 9.0 & 1.16 $\pm$ 0.26 & $-$2.69 $\pm$ 0.40 & 0.31 & 0.25 & 0.00 & 0.40  \\
 HST\_272 & 03:35:36.31 & $-$54:07:15.71 & 10.6 & 1.10 $\pm$ 0.19 & $-$2.78 $\pm$ 0.37 & 0.33 & 0.17 & 0.00 & 0.37  \\
 HST\_273 & 03:35:34.35 & $-$54:06:58.34 & 6.4 & 1.71 $\pm$ 0.37 & $-$2.49 $\pm$ 0.41 & 0.34 & 0.23 & 0.00 & 0.41  \\
 HST\_274 & 03:35:33.32 & $-$54:06:50.28 & 4.9 & 3.62 $\pm$ 0.65 & $-$2.02 $\pm$ 0.27 & 0.24 & 0.12 & 0.00 & 0.27  \\
 HST\_276 & 03:35:34.13 & $-$54:06:05.79 & 5.7 & 1.45 $\pm$ 0.42 & $-$2.59 $\pm$ 0.45 & 0.32 & 0.31 & 0.00 & 0.45  \\
 HST\_280 & 03:35:37.84 & $-$54:05:33.67 & 7.6 & 1.73 $\pm$ 0.28 & $-$2.41 $\pm$ 0.34 & 0.29 & 0.18 & 0.00 & 0.34  \\
 MP18\_000 & 03:34:18.17 & $-$54:06:21.45 & 6.5 & 1.31 $\pm$ 0.38 & $-$2.59 $\pm$ 0.45 & 0.32 & 0.32 & 0.00 & 0.45  \\
 MP18\_004 & 03:34:36.09 & $-$54:08:20.47 & 7.7 & 1.63 $\pm$ 0.29 & $-$2.41 $\pm$ 0.35 & 0.29 & 0.20 & 0.00 & 0.35  \\
 MP18\_011 & 03:34:46.83 & $-$54:04:44.62 & 4.1 & 0.45 $\pm$ 0.63 & $-$3.73 $\pm$ 0.64 & 0.30 & 0.56 & 0.00 & 0.64  \\
 MP18\_016 & 03:34:59.35 & $-$53:58:23.98 & 19.1 & 0.44 $\pm$ 0.13 & $-$3.73 $\pm$ 0.39 & 0.30 & 0.25 & 0.00 & 0.39  \\
 MP18\_025 & 03:35:13.18 & $-$54:00:33.46 & 12.8 & 6.21 $\pm$ 0.21 & $-$1.79 $\pm$ 0.22 & 0.22 & 0.05 & 0.00 & 0.23  \\
 MP18\_027 & 03:35:14.53 & $-$54:04:45.57 & 10.1 & 0.81 $\pm$ 0.21 & $-$2.99 $\pm$ 0.38 & 0.30 & 0.23 & 0.00 & 0.38  \\
 MP18\_029 & 03:35:15.45 & $-$54:04:20.48 & 10.9 & 0.79 $\pm$ 0.20 & $-$2.98 $\pm$ 0.40 & 0.32 & 0.24 & 0.00 & 0.40  \\
 MP18\_030 & 03:35:16.21 & $-$53:55:31.64 & 6.8 & 1.10 $\pm$ 0.33 & $-$2.78 $\pm$ 0.45 & 0.33 & 0.31 & 0.00 & 0.45  \\
 MP18\_038 & 03:35:24.68 & $-$54:01:19.65 & 5.2 & 1.21 $\pm$ 0.40 & $-$2.78 $\pm$ 0.50 & 0.35 & 0.36 & 0.00 & 0.50  \\
 MP18\_040 & 03:35:25.72 & $-$53:58:47.98 & 9.2 & 0.93 $\pm$ 0.27 & $-$2.98 $\pm$ 0.44 & 0.32 & 0.29 & 0.00 & 0.43  \\
 MP18\_043 & 03:35:28.60 & $-$54:10:42.04 & 5.5 & 1.73 $\pm$ 0.37 & $-$2.49 $\pm$ 0.41 & 0.34 & 0.22 & 0.00 & 0.41  \\
 MP18\_045 & 03:35:32.05 & $-$54:01:46.47 & 9.6 & 0.43 $\pm$ 0.24 & $-$3.61 $\pm$ 0.46 & 0.28 & 0.35 & 0.00 & 0.45  \\
 MP18\_056 & 03:35:41.67 & $-$53:59:03.14 & 12.0 & 4.27 $\pm$ 0.24 & $-$1.75 $\pm$ 0.22 & 0.21 & 0.07 & 0.00 & 0.22  \\
 MP18\_059 & 03:35:42.76 & $-$54:02:41.69 & 8.6 & 0.84 $\pm$ 0.25 & $-$2.99 $\pm$ 0.42 & 0.30 & 0.28 & 0.00 & 0.41  \\
 MP18\_064 & 03:35:46.61 & $-$54:04:14.98 & 3.3 & 0.89 $\pm$ 0.75 & $-$2.98 $\pm$ 0.74 & 0.32 & 0.66 & 0.00 & 0.74  \\
 MP18\_067 & 03:35:47.61 & $-$54:02:14.16 & 14.3 & 1.13 $\pm$ 0.16 & $-$2.87 $\pm$ 0.40 & 0.37 & 0.15 & 0.00 & 0.40  \\
 MP18\_076 & 03:35:54.08 & $-$54:02:19.53 & 10.7 & 3.94 $\pm$ 0.31 & $-$1.92 $\pm$ 0.24 & 0.23 & 0.07 & 0.00 & 0.24  \\
 MP18\_088 & 03:36:01.69 & $-$53:55:17.24 & 11.9 & 0.96 $\pm$ 0.17 & $-$2.98 $\pm$ 0.37 & 0.32 & 0.18 & 0.00 & 0.37  \\
 MP18\_092 & 03:36:05.22 & $-$54:01:26.11 & 10.9 & 1.11 $\pm$ 0.21 & $-$2.78 $\pm$ 0.38 & 0.33 & 0.19 & 0.00 & 0.38  \\
 MP18\_097 & 03:36:10.11 & $-$54:06:26.73 & 10.1 & 1.00 $\pm$ 0.24 & $-$2.69 $\pm$ 0.40 & 0.31 & 0.25 & 0.00 & 0.40  \\
 MP18\_098 & 03:36:10.14 & $-$53:54:46.11 & 9.4 & 1.18 $\pm$ 0.24 & $-$2.87 $\pm$ 0.42 & 0.37 & 0.20 & 0.00 & 0.42  \\
 MP18\_100 & 03:36:12.92 & $-$53:59:45.67 & 10.7 & 0.41 $\pm$ 0.23 & $-$3.61 $\pm$ 0.43 & 0.27 & 0.34 & 0.00 & 0.43  \\
 MP18\_101 & 03:36:13.42 & $-$53:59:52.11 & 14.9 & 0.56 $\pm$ 0.15 & $-$3.30 $\pm$ 0.37 & 0.30 & 0.21 & 0.00 & 0.37  \\
 MP18\_105 & 03:36:20.54 & $-$54:02:21.11 & 7.5 & 1.17 $\pm$ 0.31 & $-$2.78 $\pm$ 0.43 & 0.33 & 0.28 & 0.00 & 0.43  \\
 MP18\_108 & 03:36:26.44 & $-$54:00:57.40 & 4.9 & 1.35 $\pm$ 0.51 & $-$2.59 $\pm$ 0.53 & 0.32 & 0.42 & 0.00 & 0.53  \\
 MP18\_109 & 03:36:26.89 & $-$53:53:34.17 & 8.8 & 1.03 $\pm$ 0.29 & $-$2.78 $\pm$ 0.46 & 0.35 & 0.29 & 0.00 & 0.46  \\
 MP18\_114 & 03:36:36.55 & $-$54:03:23.83 & 4.3 & 3.11 $\pm$ 0.80 & $-$1.93 $\pm$ 0.32 & 0.23 & 0.22 & 0.00 & 0.32  \\
 MP18\_115 & 03:36:38.58 & $-$54:00:47.27 & 8.7 & 3.76 $\pm$ 0.36 & $-$1.84 $\pm$ 0.23 & 0.22 & 0.07 & 0.00 & 0.23  \\
 MP18\_119 & 03:36:43.17 & $-$53:58:15.91 & 15.4 & 0.50 $\pm$ 0.14 & $-$3.30 $\pm$ 0.37 & 0.30 & 0.22 & 0.00 & 0.37  \\
 MP18\_121 & 03:36:48.27 & $-$54:01:13.91 & 8.0 & 0.77 $\pm$ 0.30 & $-$2.98 $\pm$ 0.48 & 0.32 & 0.35 & 0.00 & 0.48  \\
 MP18\_122 & 03:36:48.93 & $-$54:00:52.51 & 7.7 & 0.75 $\pm$ 0.33 & $-$2.99 $\pm$ 0.50 & 0.30 & 0.39 & 0.00 & 0.50  \\
 MP18\_123 & 03:36:57.59 & $-$54:08:23.01 & 7.2 & 3.72 $\pm$ 0.46 & $-$2.02 $\pm$ 0.25 & 0.24 & 0.07 & 0.00 & 0.25  \\
 PL18\_004 & 03:34:20.61 & $-$54:04:34.12 & 9.9 & 1.13 $\pm$ 0.23 & $-$2.87 $\pm$ 0.43 & 0.37 & 0.21 & 0.04 & 0.43  \\
 PL18\_007 & 03:37:09.65 & $-$53:59:23.27 & 6.2 & 4.42 $\pm$ 0.52 & $-$1.83 $\pm$ 0.26 & 0.22 & 0.11 & 0.08 & 0.26  \\
 PL18\_009 & 03:34:36.71 & $-$54:06:44.93 & 32.1 & 6.65 $\pm$ 0.12 & $-$1.95 $\pm$ 0.24 & 0.24 & 0.03 & 0.00 & 0.24  \\
 PL18\_012 & 03:35:15.19 & $-$54:08:42.99 & 25.7 & 3.01 $\pm$ 0.12 & $-$2.63 $\pm$ 0.29 & 0.29 & 0.05 & 0.00 & 0.29  \\
 PL18\_013 & 03:34:47.95 & $-$54:05:24.99 & 75.4 & 3.13 $\pm$ 0.04 & $-$2.89 $\pm$ 0.23 & 0.23 & 0.00 & 0.00 & 0.23  \\
 PL18\_016 & 03:35:48.06 & $-$54:03:49.79 & 25.0 & 6.36 $\pm$ 0.17 & $-$2.05 $\pm$ 0.24 & 0.24 & 0.04 & 0.00 & 0.24  \\
 PL18\_018 & 03:36:43.06 & $-$53:53:53.55 & 24.1 & 1.69 $\pm$ 0.09 & $-$2.67 $\pm$ 0.38 & 0.38 & 0.05 & 0.00 & 0.38  \\
 HST\_013 & 03:35:47.28 & $-$54:00:46.72 & 9.0  & 0.76 & $<$ $-$3.09 & \dots & \dots & \dots & \dots \\
 HST\_014 & 03:35:41.22 & $-$54:00:30.62 & 4.8  & 1.46 & $<$ $-$2.59 & \dots & \dots & \dots & \dots \\
 HST\_027 & 03:36:11.72 & $-$54:01:12.69 & 7.0  & 1.00 & $<$ $-$2.60 & \dots & \dots & \dots & \dots \\
 HST\_041 & 03:36:08.99 & $-$54:00:13.78 & 7.4  & 0.97 & $<$ $-$3.09 & \dots & \dots & \dots & \dots \\
 HST\_046 & 03:36:16.89 & $-$54:01:27.43 & 4.5  & 1.84 & $<$ $-$2.48 & \dots & \dots & \dots & \dots \\
 HST\_051 & 03:36:22.80 & $-$54:00:57.18 & 3.2  & 2.31 & $<$ $-$2.06 & \dots & \dots & \dots & \dots \\
 HST\_052 & 03:36:36.36 & $-$54:00:50.11 & 4.6  & 1.61 & $<$ $-$2.49 & \dots & \dots & \dots & \dots \\
 HST\_060 & 03:36:28.00 & $-$54:00:11.31 & 5.6  & 1.23 & $<$ $-$2.78 & \dots & \dots & \dots & \dots \\
 HST\_082 & 03:34:57.97 & $-$54:02:39.02 & 5.0  & 1.43 & $<$ $-$2.58 & \dots & \dots & \dots & \dots \\
 HST\_084 & 03:34:50.94 & $-$54:01:53.82 & 4.9  & 1.33 & $<$ $-$2.58 & \dots & \dots & \dots & \dots \\
 HST\_108 & 03:35:11.81 & $-$54:03:27.17 & 7.3  & 1.04 & $<$ $-$2.78 & \dots & \dots & \dots & \dots \\
 HST\_130 & 03:35:18.05 & $-$54:01:59.15 & 8.0  & 0.88 & $<$ $-$2.98 & \dots & \dots & \dots & \dots \\
 HST\_166 & 03:35:50.93 & $-$54:03:05.35 & 9.4  & 0.73 & $<$ $-$3.30 & \dots & \dots & \dots & \dots \\
 HST\_175 & 03:35:31.54 & $-$54:02:30.57 & 7.2  & 1.00 & $<$ $-$2.98 & \dots & \dots & \dots & \dots \\
 HST\_184 & 03:36:10.76 & $-$54:04:32.86 & 3.4  & 2.17 & $<$ $-$2.21 & \dots & \dots & \dots & \dots \\
 HST\_195 & 03:35:57.73 & $-$54:02:57.31 & 6.8  & 1.01 & $<$ $-$2.78 & \dots & \dots & \dots & \dots \\
 HST\_196 & 03:35:55.70 & $-$54:02:56.74 & 4.7  & 1.56 & $<$ $-$2.49 & \dots & \dots & \dots & \dots \\
 HST\_206 & 03:36:07.58 & $-$54:02:07.22 & 3.1  & 2.25 & $<$ $-$2.12 & \dots & \dots & \dots & \dots \\
 HST\_208 & 03:36:14.29 & $-$54:01:52.83 & 3.6  & 2.03 & $<$ $-$2.22 & \dots & \dots & \dots & \dots \\
 HST\_212 & 03:36:09.08 & $-$54:01:39.50 & 3.5  & 2.08 & $<$ $-$2.22 & \dots & \dots & \dots & \dots \\
 HST\_245 & 03:34:59.64 & $-$54:04:44.70 & 10.9  & 0.69 & $<$ $-$3.30 & \dots & \dots & \dots & \dots \\
 HST\_246 & 03:35:11.29 & $-$54:07:57.13 & 4.7  & 1.38 & $<$ $-$2.67 & \dots & \dots & \dots & \dots \\
 HST\_247 & 03:35:09.58 & $-$54:07:41.51 & 4.5  & 1.43 & $<$ $-$2.67 & \dots & \dots & \dots & \dots \\
 HST\_249 & 03:35:11.61 & $-$54:07:26.90 & 6.2  & 1.07 & $<$ $-$2.78 & \dots & \dots & \dots & \dots \\
 MP18\_002 & 03:34:19.95 & $-$54:05:12.11 & 4.2  & 1.95 & $<$ $-$2.23 & \dots & \dots & \dots & \dots \\
 MP18\_010 & 03:34:41.34 & $-$54:03:31.83 & 9.6  & 0.71 & $<$ $-$3.30 & \dots & \dots & \dots & \dots \\
 MP18\_017 & 03:35:00.43 & $-$54:04:00.37 & 4.2  & 1.54 & $<$ $-$2.49 & \dots & \dots & \dots & \dots \\
 MP18\_018 & 03:35:03.11 & $-$53:57:41.20 & 13.4  & 0.53 & $<$ $-$3.30 & \dots & \dots & \dots & \dots \\
 MP18\_032 & 03:35:21.69 & $-$54:07:24.92 & 3.4  & 2.21 & $<$ $-$2.22 & \dots & \dots & \dots & \dots \\
 MP18\_046 & 03:35:32.19 & $-$54:02:14.17 & 3.0  & 2.51 & $<$ $-$1.97 & \dots & \dots & \dots & \dots \\
 MP18\_051 & 03:35:38.28 & $-$54:08:44.56 & 5.0  & 1.42 & $<$ $-$2.59 & \dots & \dots & \dots & \dots \\
 MP18\_054 & 03:35:40.77 & $-$54:11:05.57 & 5.0  & 1.22 & $<$ $-$2.78 & \dots & \dots & \dots & \dots \\
 MP18\_063 & 03:35:46.36 & $-$54:10:47.49 & 6.6  & 1.06 & $<$ $-$2.87 & \dots & \dots & \dots & \dots \\
 MP18\_068 & 03:35:48.24 & $-$53:59:45.28 & 8.8  & 0.65 & $<$ $-$3.30 & \dots & \dots & \dots & \dots \\
 MP18\_074 & 03:35:53.06 & $-$53:56:08.24 & 11.8  & 0.55 & $<$ $-$3.19 & \dots & \dots & \dots & \dots \\
 MP18\_078 & 03:35:55.03 & $-$54:02:32.48 & 4.9  & 1.50 & $<$ $-$2.49 & \dots & \dots & \dots & \dots \\
 MP18\_083 & 03:35:57.48 & $-$53:55:56.77 & 9.8  & 0.73 & $<$ $-$3.19 & \dots & \dots & \dots & \dots \\
 MP18\_099 & 03:36:12.77 & $-$54:04:18.36 & 4.7  & 1.75 & $<$ $-$2.49 & \dots & \dots & \dots & \dots
\end{longtable}
\tablecomments{\textcolor{vibrantpurple}{\textbf{*}} EW for upper limits are $3\sigma_\text{EW}$.}
\twocolumngrid 
\hphantom{1pt}
\clearpage
\section{Posterior Distributions}
\label{appendix}
The posterior distributions of the model parameters for the Leaky Box, Gaussian, Leaky Box + Gaussian Mixture Model, and Gaussian + Gaussian Mixture Model. Each model is sampled for two datasets: $e_{\feh, \text{total}}$ and $e_\text{[Fe/H], spec}$. The total uncertainties, $e_\feh$, are larger due to the systematics in the metallicity calibration, $e_\text{[Fe/H], calib}$, that are added in quadrature to the spectroscopic uncertainties, $e_\text{[Fe/H], spec}$. For unconstrained parameters, we report the 90th percentile upper limits in Table \ref{tab:bimodal_params}. 
\begin{figure}[ht!]
    \centering
    \includegraphics[width=0.7\linewidth]{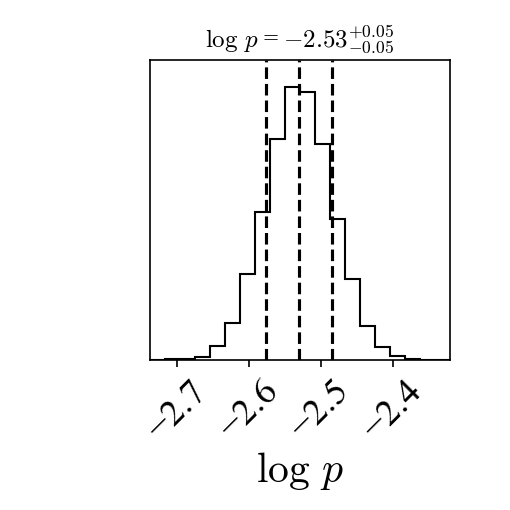}
    \caption{Posterior distribution of the Leaky Box model parameter, $p$. Inputs are the metallicities including upper limits, \feh, and the total uncertainty, $e_{\feh,\text{total}} = \sqrt{e_{\feh, \text{spec}}^2 + e_{\feh, \text{calib}}^2}$.}
    \label{fig:enter-label1}
\end{figure}
\begin{figure}
    \centering
    \includegraphics[width=0.7\linewidth]{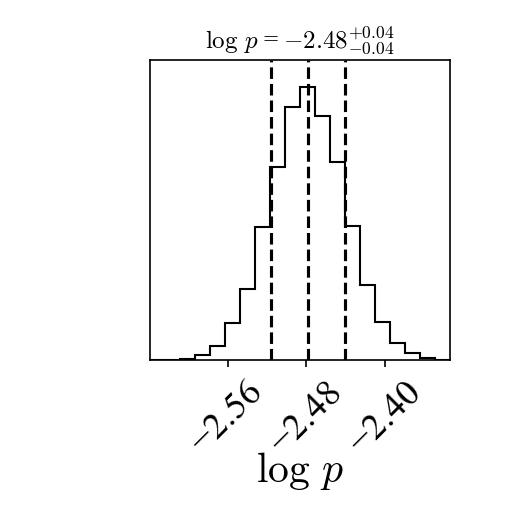}
    \caption{Posterior distribution of the Leaky Box model parameter, $p$. Inputs are the metallicities including upper limits, \feh, and the spectroscopic uncertainty, $e_{\feh, \text{spec}}$.}
    \label{fig:enter-label2}
\end{figure}
\\
\\

\begin{figure}[ht!]
    \centering
    \includegraphics[width=\linewidth]{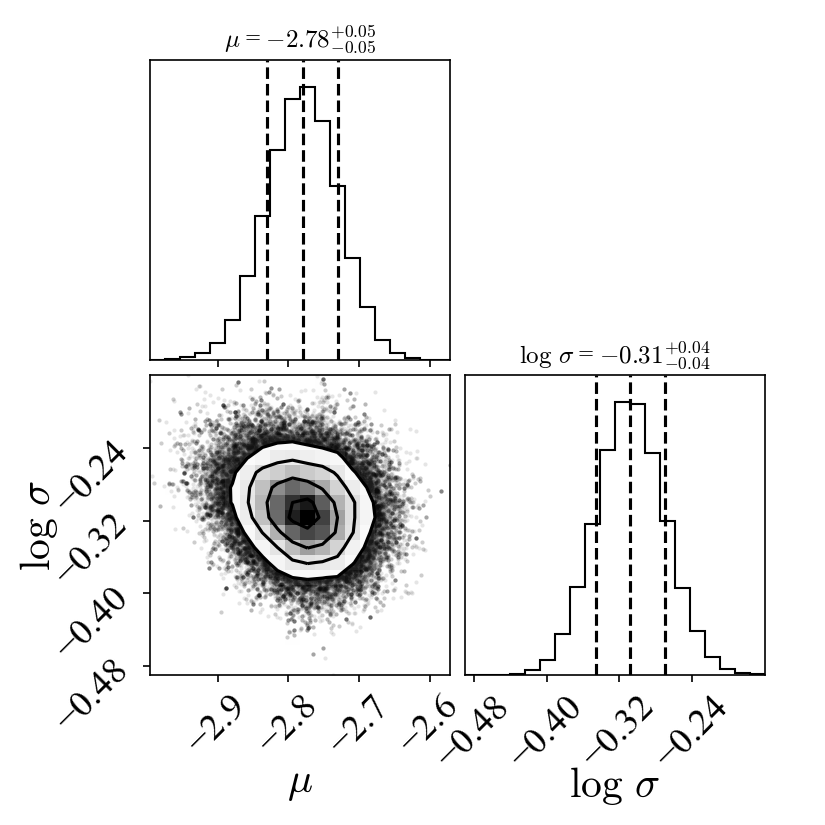}
    \caption{Posterior distribution of Gaussian model parameters: $\mu \ \text{and} \  \sigma$. Inputs = \feh, $e_{\feh, \text{total}}$.}
    \label{fig:enter-label3}
\end{figure}

\begin{figure}
    \centering
    \includegraphics[width=\linewidth]{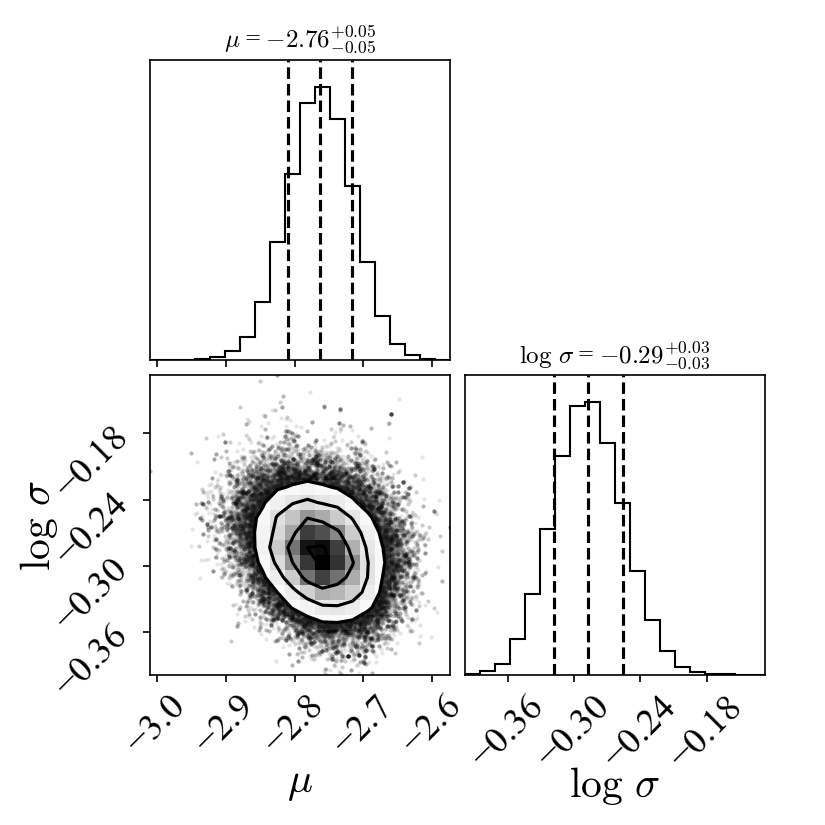}
    \caption{Posterior distribution of Gaussian model parameters: $\mu  \ \text{and} \ \sigma$. Inputs = \feh, $e_{\feh, \text{spec}}$.}
    \label{fig:enter-label4}
\end{figure}

\begin{figure}
    \centering
    \includegraphics[width=\linewidth]{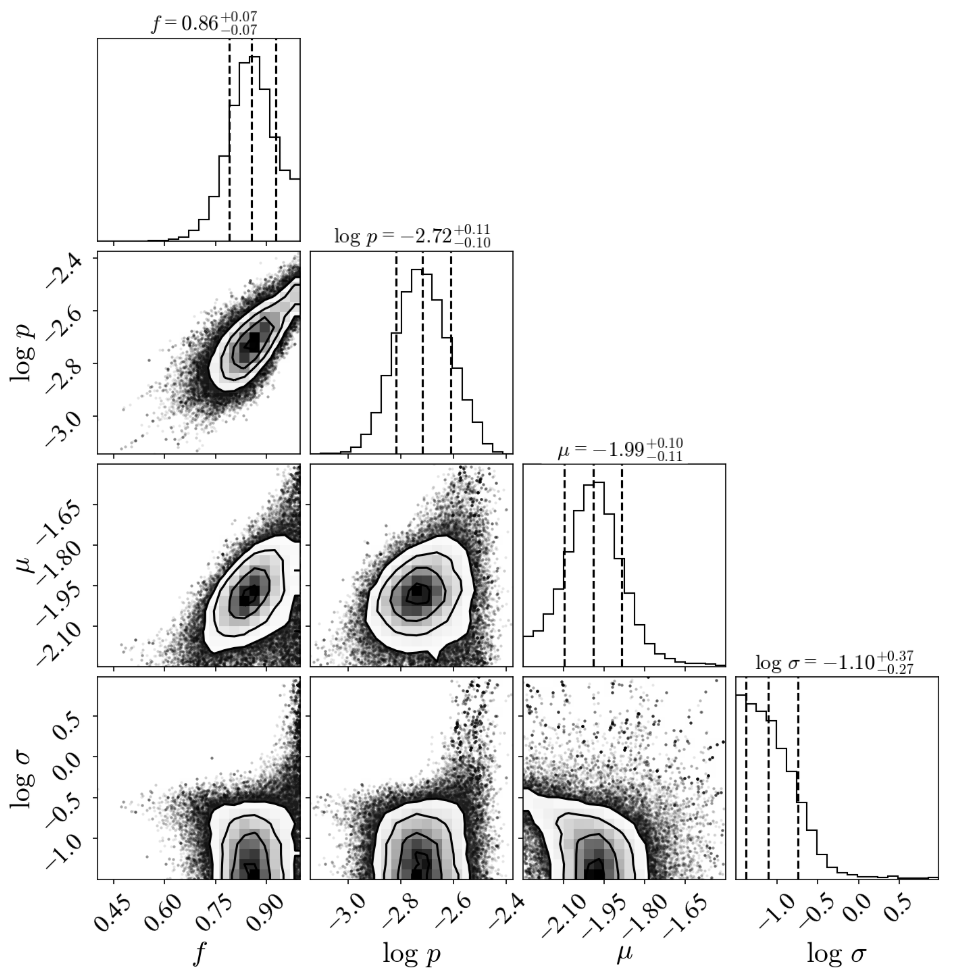}
    \caption{Posterior distribution of Leaky Box +  Gaussian model parameters: $ f, p, \mu,  \ \text{and} \ \sigma$. Inputs = \feh, $e_{\feh, \text{total}}$.}
    \label{fig:enter-label5}
\end{figure}

\begin{figure}
    \centering
    \includegraphics[width=\linewidth]{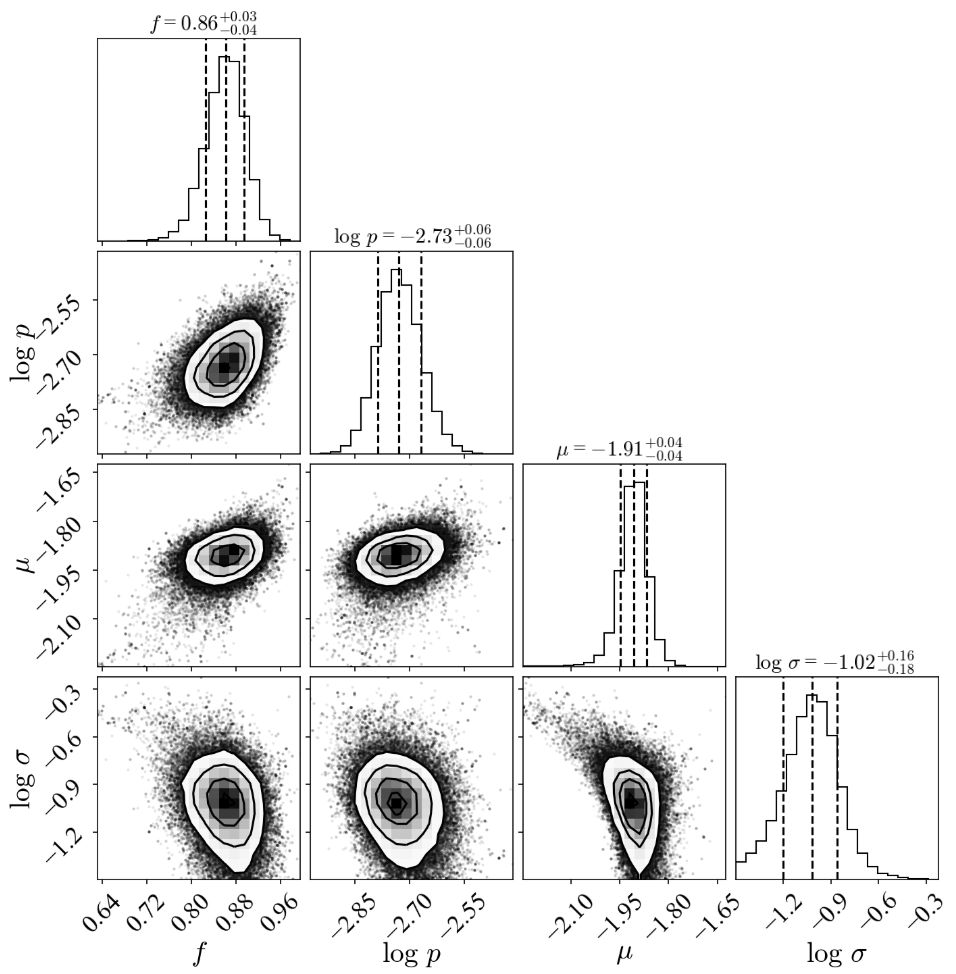}
    \caption{Posterior distribution of Leaky Box + Gaussian model parameters: $ f, p, \mu,  \ \text{and} \ \sigma$.  Inputs = \feh, $e_{\feh, \text{spec}}$.}
    \label{fig:enter-label6}
\end{figure}

\begin{figure}
    \centering
    \includegraphics[width=\linewidth]{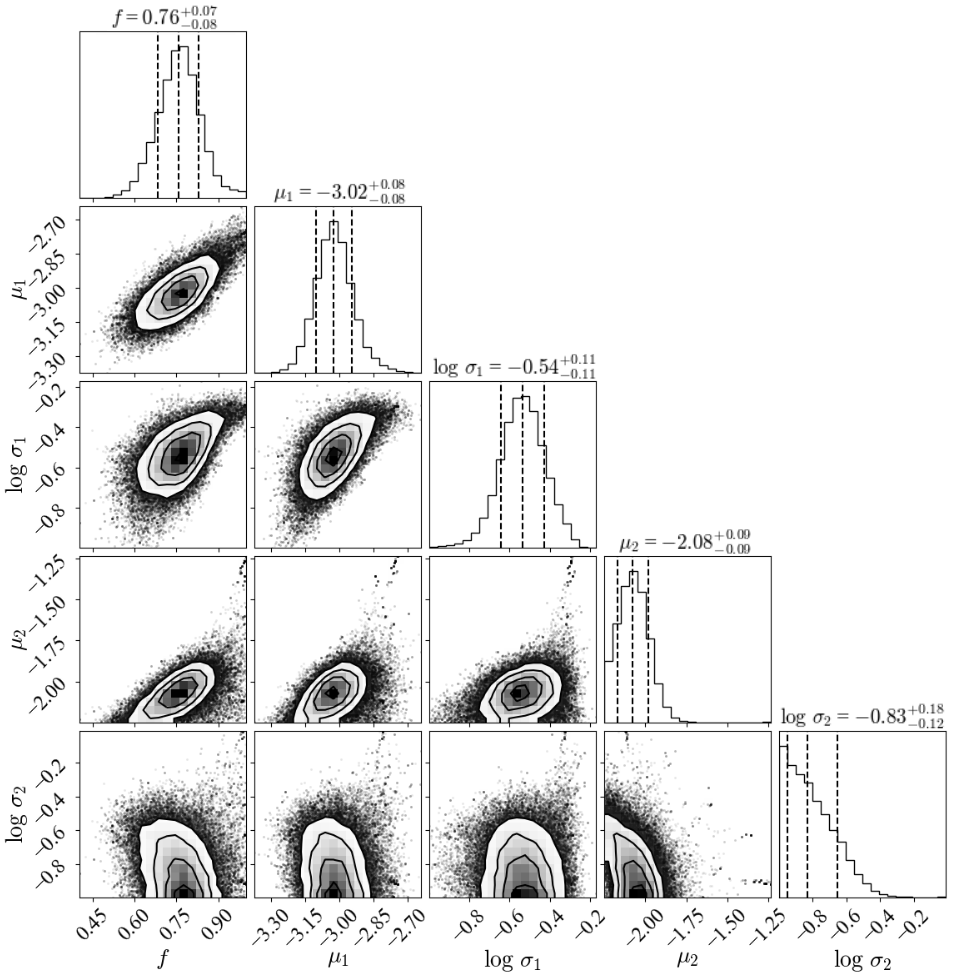}
    \caption{Posterior distribution of Gaussian + Gaussian model parameters:  $f, \mu_1, \sigma_1, \mu_2, \ \text{and} \  \sigma_2$. Inputs = \feh, $e_{\feh, \text{total}}$.}
    \label{fig:enter-label7}
\end{figure}

\begin{figure}
    \centering
    \includegraphics[width=\linewidth]{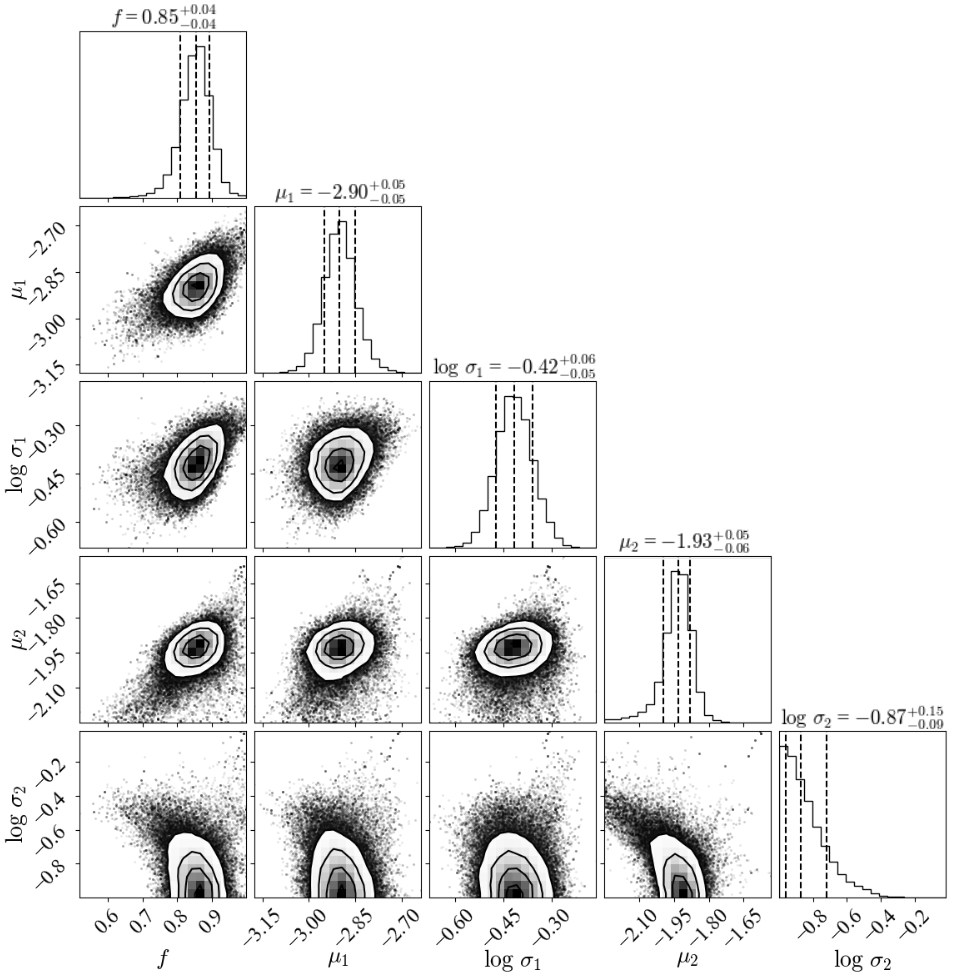}
    \caption{Posterior distribution of Gaussian + Gaussian model parameters:  $f, \mu_1, \sigma_1, \mu_2, \ \text{and} \  \sigma_2$.  Inputs = \feh, $e_{\feh, \text{spec}}$.}
    \label{fig:enter-label8}
\end{figure}
\vspace{10cm}

\end{document}